\documentclass[journal]{IEEEtran}
\usepackage{amsmath,amssymb,amsthm}
\usepackage{graphicx}
\usepackage{microtype} % micro-typography, slightly tighter paragraphs
\usepackage{booktabs}
\usepackage{url}
\usepackage[colorlinks=true,allcolors=blue]{hyperref}
\usepackage{array}
\usepackage{algorithm}
\usepackage{algpseudocode}
\algrenewcommand\algorithmicrequire{\textbf{Input:}}
\algrenewcommand\algorithmicensure{\textbf{Output:}}
\algdef{SE}[DOWHILE]{Do}{doWhile}{\algorithmicdo}[1]{\algorithmicwhile\ #1}
\makeatletter
\newcommand{\algml}[1]{\parbox[t]{\dimexpr\linewidth-\ALG@thistlm\relax}{\raggedright\strut #1\strut}}
\makeatother
\newlength{\alglabelwidth}
\newcommand{\algmlin}[1]{\settowidth{\alglabelwidth}{\algorithmicrequire\ }\parbox[t]{\dimexpr\linewidth-\alglabelwidth\relax}{\raggedright\strut #1\strut}}

\usepackage{dsfont}
\usepackage{tikz}
\usetikzlibrary{arrows.meta, positioning, fit, backgrounds}
\usepackage{acronym}
\usepackage{comment}

\makeatletter
\g@addto@macro\normalsize{\setlength\abovedisplayskip{4pt plus 2pt minus 1pt}\setlength\belowdisplayskip{4pt plus 2pt minus 1pt}\setlength\abovedisplayshortskip{2pt plus 1pt}\setlength\belowdisplayshortskip{2pt plus 1pt}}
\makeatother
\newcommand{\Ex}{\mathbb{E}}

\newcommand{\diag}{\mathrm{diag}}

\newcommand{\range}{\mathrm{range}}
\newcommand{\Herm}{\mathsf{H}}
\newcommand{\T}{\mathsf{T}}
\newcommand{\Vb}{\overline{\mathbf{V}}}
\newcommand{\Ub}{\overline{\mathbf{U}}}
\newcommand{\Sb}{\overline{\boldsymbol{\Sigma}}}
\newcommand{\Yb}{\mathbf{Y}}
\newcommand{\Wb}{\mathbf{W}}
\newcommand{\Phib}{\boldsymbol{\Phi}}
\DeclareMathOperator{\tr}{tr}
\usepackage{units}

\newtheorem{lemma}{Lemma}

\theoremstyle{remark}
\newtheorem{remark}{Remark}
\usepackage{acronym}

\acrodef{RIS}{reconfigurable intelligent surface}
\acrodef{ISAC}{integrated sensing and communication}
\acrodef{SIM}{stacked intelligent metasurface}
\acrodef{FIM}{fluid intelligent metasurface}
\acrodef{MIMO}{multiple-input multiple-output}
\acrodef{MiLAC}{microwave linear analog computer}
\acrodef{RF}{radio frequency}
\acrodef{AWGN}{additive white Gaussian noise}
\acrodef{SVD}{singular value decomposition}
\acrodef{SNR}{signal-to-noise-ratio}
\acrodef{i.i.d.}{independent and identically distributed}
\acrodef{MC}{Monte Carlo}
\acrodef{SINR}{signal-to-interference-plus-noise-ratio}
\acrodef{LoS}{line-of-sight}
\acrodef{LMMSE}{linear minimum mean-square error}
\acrodef{MU-MISO}{multi-user multiple-input single-output}
\acrodef{DBF}{digital beamforming}

\begin{document}

\title{\huge Intrinsic Impedance Mismatch in MiLAC-Aided Communications: Rate Analysis and Robust Design \vspace{-0.15cm}}  
\author{Ahmed Magbool,~\IEEEmembership{Member,~IEEE,} and Marco Di Renzo,~\IEEEmembership{Fellow,~IEEE}\thanks{Ahmed Magbool and Marco Di Renzo are with the Centre for Telecommunications Research, Department of Engineering, King's College London, WC2R 2LS London, United Kingdom (e-mail: ahmed.1.magbool@kcl.ac.uk; marco.di\_renzo@kcl.ac.uk). Marco Di Renzo is also with CNRS and CentraleSup\'elec, Institute of Electronics and Digital Technologies (IETR), Avenue de la Boulaie, 35576 Cesson-S\'evign\'e, France (e-mail: marco.direnzo@centralesupelec.fr). \par
%This work was supported in part by the European Research Council (ERC) under the European Union's Horizon Europe Programme WePhICom (agreement number 101225119), as well as by the European Union through the Horizon Europe project COVER under grant agreement number 101086228, the Horizon Europe project UNITE under grant agreement number 101129618, the Horizon Europe project INSTINCT under grant agreement number 101139161, and the Horizon Europe project TWIN6G under grant agreement number 101182794, as well as by the Agence Nationale de la Recherche (ANR) through the France 2030 project ANR-PEPR Networks of the Future under grant agreement NF-YACARI 22-PEFT-0005, and by the CHIST-ERA project PASSIONATE under grant agreements CHIST-ERA-22-WAI-04 and ANR-23-CHR4-0003-01. Also, the work of M. Di Renzo was supported in part by the Engineering and Physical Sciences Research Council (EPSRC), part of UK Research and Innovation, and the UK Department of Science, Innovation and Technology through the CHEDDAR Telecom Hub under grant EP/Y037421/1, through the HASC Telecom Hub under grant EP/Y037197/1, and through the TITAN Telecom Hub under grant EP/Y037243/1.
} \vspace{-1.1cm}}

\maketitle

\begin{abstract}
Microwave linear analog computers (MiLACs) are reconfigurable multiport networks that perform linear operations in the analog domain, making them promising enablers of beamforming in large-scale multiple-input multiple-output (MIMO) systems. However, existing MiLAC designs assume that the tunable admittances realize their prescribed values exactly, whereas practical reactive components are subject to manufacturing tolerances. This paper develops a framework for analyzing and designing MiLAC-aided MIMO systems under intrinsic impedance mismatch, based on a uniform statistical model of the admittance errors. Using perturbation analysis, we derive closed-form approximations of the expected sum rate and characterize its key properties. We show that, unlike in the error-free case, the expected rate saturates at high transmit power, and we derive the rate ceiling in closed form. We also show that, for a fixed ratio of streams to antennas, doubling the number of antennas is as detrimental as increasing the tolerance by a factor of $\sqrt{2}$. We further show that errors in the receive MiLAC are generally more detrimental than errors in the transmit MiLAC. We then propose a robust design that jointly optimizes the two scattering matrices and the power allocation using a monotone accelerated projected gradient ascent method. Numerical results demonstrate that intrinsic impedance mismatch can cause substantial rate degradations. For example, with 32 antennas and a transmit power of 30 dBm, the capacity-achieving design suffers a rate degradation of approximately 49\% under a normalized tolerance of 10\%, whereas the proposed robust design recovers approximately 44\% of this degradation.
\end{abstract}


\vspace{-0.3cm}
\section{Introduction}
Future wireless networks are expected to support high spectral efficiency using increasingly large antenna arrays, with massive \ac{MIMO} being a key enabling technology. However, this comes at the cost of increased implementation complexity and power consumption as the array size grows, since each antenna requires a dedicated \ac{RF} chain. Consequently, both hardware cost and front-end power consumption scale with the number of antennas, making fully digital architectures increasingly challenging to implement in systems with large arrays~\cite{2014LarssonMassive}.

Several architectures have been proposed to address this issue~\cite{2025YouNext}, including hybrid analog-digital beamforming~\cite{2016SohrabiHybrid}, intelligent metasurfaces~\cite{2025MagboolSurvey,2026MagboolBeyond}, and movable antennas~\cite{2026ZhuTutorial}. More recently, \ac{MiLAC} has been proposed to perform signal processing operations directly in the analog domain~\cite{2025NeriniAnalog}. A \ac{MiLAC} is a reconfigurable multiport microwave network whose tunable admittances can be configured to implement linear transformations directly in the analog domain~\cite{2025NeriniAnalog1}. By placing a \ac{MiLAC} between the \ac{RF} chains and the antennas, beamforming is performed by the circuit itself, so that signal processing takes place at the speed of electromagnetic wave propagation and the number of \ac{RF} chains is reduced to the number of data streams.

The work in~\cite{2025NeriniAnalog} demonstrated that \acp{MiLAC} can perform operations such as matrix inversion at a fraction of the arithmetic cost of digital processing. These capabilities have motivated the use of \acp{MiLAC} for beamforming in large antenna arrays~\cite{2025NeriniAnalog1}. In this context,~\cite{2025neriniCapacity} showed that a lossless and reciprocal network achieves the capacity of a point-to-point \ac{MIMO} system obtained with \ac{DBF}. In \ac{MU-MISO} systems, \ac{MiLAC}-based beamforming generally achieves a lower rate than fully \ac{DBF}, but with substantially lower power consumption~\cite{2026WuMicrowave}. Recent studies have further explored the energy efficiency of \ac{MiLAC}-enabled communications~\cite{2026MagboolEnergy,2026ZhangQuantization}, \ac{MiLAC} architectures with reduced circuit complexity~\cite{2026NeriniMIMO,2026ZhangBeamforming}, general impedance matching~\cite{2026XiongGeneral}, sensing applications~\cite{2026LiuMicrowave,2026ZhangHow}, simultaneous active and passive beamforming~\cite{2026NeriniMicrowave}, and wideband communications~\cite{2026PengHybrid}.

Nevertheless, existing studies assume that the tunable components of a \ac{MiLAC} realize their prescribed admittances exactly. In a lossless \ac{MiLAC}, these components are purely reactive and can be implemented using tunable capacitors, such as varactor diodes~\cite{LiTutorialBDRIS,LiWidebandBDRIS,TapiePrototype}, whose capacitances are subject to device variations and manufacturing tolerances~\cite{SkyworksSMV1405}. The realized admittances therefore deviate from the designed ones, which we refer to as \emph{intrinsic impedance mismatch}. \acp{MiLAC} are particularly sensitive to this mismatch for a number of reasons. The precoder and the combiner are blocks of the scattering matrix, which is obtained from the admittance matrix through a matrix inversion, so that a single component deviation perturbs all their entries. In addition, the number of tunable components, and hence of error sources, grows quadratically with the number of ports. Furthermore, the capacity-achieving configuration of~\cite{2025neriniCapacity} decomposes the \ac{MIMO} channel into interference-free parallel streams only if the admittances are realized accurately. Under mismatch, the realized beamformers no longer diagonalize the channel, which introduces inter-stream interference and alters the stream gains and the collected noise. The rate degradation under realistic component tolerances, how it scales with transmit power and array size, and whether it can be mitigated through appropriate design remain open questions.

Motivated by these open questions, we develop a framework for the analysis and design of \ac{MiLAC}-aided \ac{MIMO} systems under intrinsic impedance mismatch. The main contributions of this paper are summarized as follows:
\begin{itemize}
    \item We develop a first-order perturbation analysis of the Cayley transform to characterize how bounded uniform admittance errors affect the realized precoder and combiner. We further derive a closed-form approximation of the expected sum rate achieved by the capacity-achieving design under intrinsic impedance mismatch.
    \item Based on the derived expressions, we characterize key properties of the expected sum rate under intrinsic impedance mismatch. We show that, unlike in the error-free case, the sum rate saturates at high transmit powers, with the rate ceiling derived in closed form. We also quantify the rate degradation as a function of the number of antennas and show that, for a fixed ratio of streams to antennas, doubling the number of antennas is as detrimental as increasing the tolerance by a factor of $\sqrt{2}$. We further show that errors in the receive \ac{MiLAC} are generally more detrimental than those in the transmit \ac{MiLAC}.
    \item We formulate a robust design problem that maximizes the expected sum rate under intrinsic impedance mismatch over the transmit and receive scattering matrices and the power allocation, subject to the lossless and reciprocal constraints of the \acp{MiLAC} and the transmit power budget. To solve this non-convex problem, for which no closed-form solution is available, we propose a monotone accelerated projected gradient ascent method.
    \item We perform extensive numerical simulations to validate the analytical results against \ac{MC} simulations, quantify the gains achieved by the robust design, and compare the performance of the \ac{MiLAC} and \ac{DBF} under different hardware impairments. The results show that intrinsic impedance mismatch can cause substantial rate degradations. For instance, with $32$ antennas, a transmit power of $\unit[30]{dBm}$, and a normalized tolerance of $10\%$, the capacity-achieving design under mismatch suffers a rate degradation of approximately $49\%$, while the robust design recovers approximately $44\%$ of this degradation.
\end{itemize}

The rest of the paper is organized as follows. Section~\ref{sec:system} presents the system model and the capacity-achieving configuration. Section~\ref{sec:int} models the admittance errors, derives the expected sum rate under intrinsic impedance mismatch, and discusses its key properties. Section~\ref{sec:robust} presents the robust design. Section~\ref{sec:results} presents numerical simulations that validate the analysis, evaluate the robust design, and compare the \ac{MiLAC} with \ac{DBF}, and Section~\ref{sec:conclusion} concludes the paper.

\textit{Notation:} Bold lowercase and uppercase letters denote vectors and matrices. $[\mathbf{A}]_{m,n}$, $[\mathbf{A}]_{m,:}$, and $[\mathbf{A}]_{:,n}$ denote the $(m,n)$-th entry, the $m$-th row, and the $n$-th column of $\mathbf{A}$, and, for a block matrix, $[\mathbf{A}]_{ab}$ and $[\mathbf{A}]_{a,:}$ denote its $(a,b)$-th block and its $a$-th block row. $(\cdot)^*$, $(\cdot)^\T$, $(\cdot)^\Herm$, $(\cdot)^{-1}$, and $(\cdot)^{1/2}$ denote the complex conjugate, the transpose, the conjugate transpose, the inverse, and the principal square root. $\mathrm{Re}\{\cdot\}$, $\mathrm{Im}\{\cdot\}$, and $|\cdot|$ denote the real part, the imaginary part, and the magnitude. $\|\cdot\|_2$ denotes the Euclidean norm of a vector and the spectral norm of a matrix, $\|\cdot\|_\mathrm{F}$ the Frobenius norm, and $\tr\{\cdot\}$ the trace. $\mathbf{I}_a$ denotes the $a\times a$ identity matrix, and $\mathbf{0}_{a\times b}$ and $\mathbf{1}_{a\times b}$ denote the $a\times b$ all-zeros and all-ones matrices, where $\mathbf{0}_a$ and $\mathbf{1}_a$ denote the corresponding $a\times 1$ vectors. $\diag(a_1,\dots,a_N)$ and $\diag(\mathbf{a})$ denote the diagonal matrices whose diagonal entries are $a_1,\dots,a_N$ and the entries of $\mathbf{a}$, $\odot$ denotes the element-wise product, and $\mathbf{r}(\mathbf{A})$ and $\mathbf{c}(\mathbf{A})$ denote the vectors collecting the squared Euclidean norms of the rows and of the columns of $\mathbf{A}$. $\ker(\cdot)$ and $\range(\cdot)$ denote the null space and the range of a matrix, and $(\cdot)^\perp$ the orthogonal complement. $\Ex[\cdot]$ denotes the expectation, $\mathcal{CN}(\mathbf{a},\mathbf{B})$ denotes the circularly symmetric complex Gaussian distribution with mean $\mathbf{a}$ and covariance matrix $\mathbf{B}$, and $\mathcal{U}[a,b]$ the uniform distribution over $[a,b]$. For a real-valued function $f$ of a complex matrix $\mathbf{A}$, $\partial f/\partial\mathbf{A}^*$ denotes the Wirtinger derivative and $\nabla_{\mathbf{A}}f$ the gradient. Finally, $\mathds{1}_{\mathcal{E}}$ equals one if $\mathcal{E}$ holds and zero otherwise, $[x]^+\triangleq\max(x,0)$, and $\mathcal{O}(\cdot)$ denotes the big-O notation for the complexity.

% =====================================================================
\vspace{-0.3cm}
\section{System Model}
\label{sec:system}

\begin{figure} 
         \centering
\includegraphics[width=0.55\linewidth]{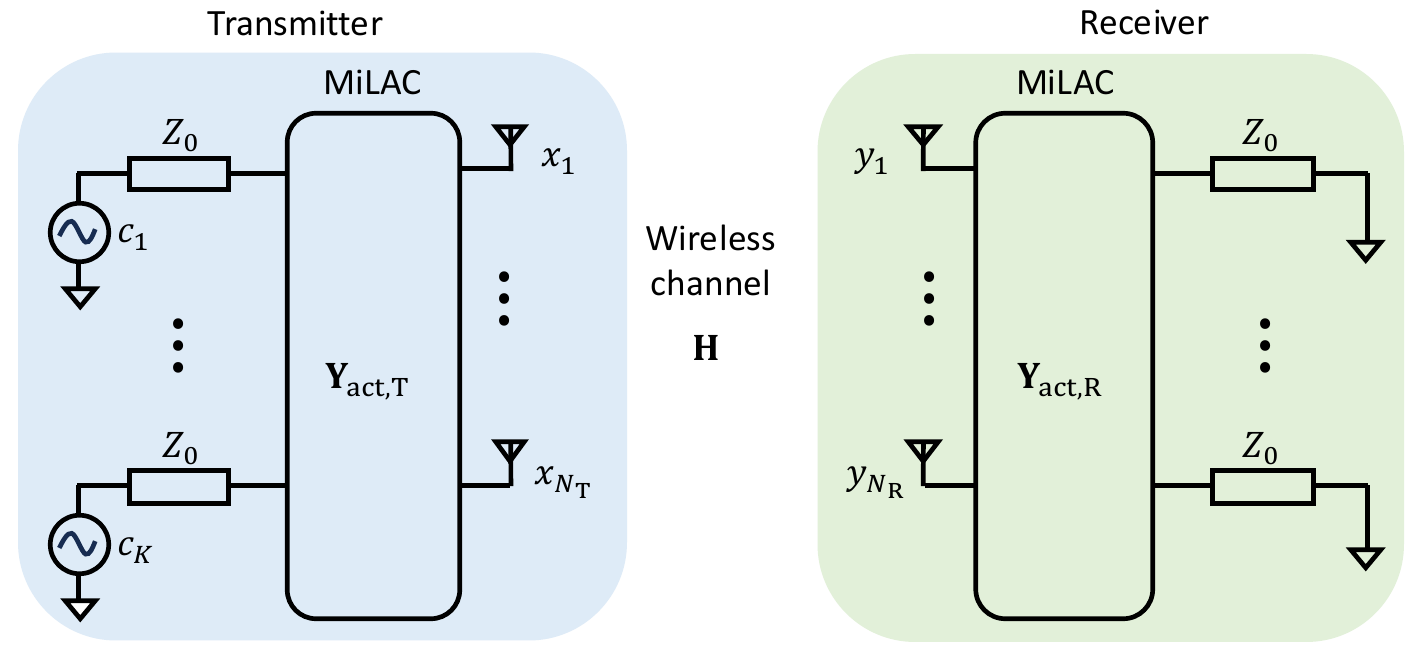}
       \vspace{-0.3cm}
       \caption{System model in which both the transmitter and receiver employ \ac{MiLAC} networks for beamforming.}
       \vspace{-0.2cm}
        \label{fig:system_model}
\end{figure}

As shown in Fig.~\ref{fig:system_model}, we consider a \ac{MIMO} system in which both the transmitter and receiver employ \ac{MiLAC} networks for beamforming. The system comprises \(N_\mathrm{T}\) transmit antennas and \(N_\mathrm{R}\) receive antennas to transmit \(N_\mathrm{S}\leq\min(N_\mathrm{T},N_\mathrm{R})\) independent data streams\footnote{We consider multi-stream transmission ($N_\mathrm{S}\geq 2$), which is of practical interest for \ac{MiLAC}-aided \ac{MIMO} systems. Although the single-stream case ($N_\mathrm{S}=1$) follows directly from the derived expressions, its properties differ fundamentally and are omitted for brevity.}. Accordingly, the transmitter-side \ac{MiLAC} network has \(M_\mathrm{T}\triangleq N_\mathrm{S}+N_\mathrm{T}\) ports, with the first \(N_\mathrm{S}\) connected to the \ac{RF} chains and the remaining \(N_\mathrm{T}\) connected to the transmit antennas. Similarly, the receiver-side \ac{MiLAC} network comprises $M_\mathrm{R}\triangleq N_\mathrm{S}+N_\mathrm{R}$ ports.

\vspace{-0.3cm}

\subsection{Transmit and Receive Signal Models}
\label{subsec:signal-model}
The signal at the transmit \ac{RF} chains is $\mathbf{c}=\sqrt{P_\mathrm{T}}\,\mathbf{P}^{1/2}\mathbf{s},$ where $P_\mathrm{T}$ is the total transmit power budget, $\mathbf{P}\triangleq\diag(p_1,\dots,p_{N_\mathrm{S}})$ with $p_k$ denoting the power fraction allocated to the $k$-th stream and $\sum_kp_k=1$, and $\mathbf{s} = [s_1,\dots,s_{N_\mathrm{S}}]^\T\sim\mathcal{CN}(\mathbf{0},\mathbf{I}_{N_\mathrm{S}})$ is the data vector.

The transmitter applies the precoder $\mathbf{F}\in\mathbb{C}^{N_\mathrm{T}\times N_\mathrm{S}}$ through its \ac{MiLAC}, and the receiver applies the combiner $\mathbf{G}\in\mathbb{C}^{N_\mathrm{S}\times N_\mathrm{R}}$ through its \ac{MiLAC}, yielding a post-processed received signal in the form $\mathbf{q}=\sqrt{P_\mathrm{T}}\,\mathbf{G}\mathbf{H}\mathbf{F}\mathbf{P}^{1/2}\mathbf{s}+\mathbf{G}\mathbf{n},$ where $\mathbf{H}\in\mathbb{C}^{N_\mathrm{R}\times N_\mathrm{T}}$ is the channel matrix and $\mathbf{n}\sim\mathcal{CN}(\mathbf{0},\varpi^2\mathbf{I}_{N_\mathrm{R}})$ is the \ac{AWGN} at the receiver. 

Denoting the $k$-th column of $\mathbf{F}$ by $\mathbf{f}_k$ and the $k$-th row of $\mathbf{G}$ by $\mathbf{g}_k^{\mathsf{T}}$, the $k$-th entry of $\mathbf{q}$, which is used to detect the $k$-th data stream, consists of the desired signal $\sqrt{P_\mathrm{T}p_k}\,\mathbf{g}_k^{\mathsf{T}}\mathbf{H}\mathbf{f}_k s_k$, the inter-stream interference $\sum_{j\neq k}\sqrt{P_\mathrm{T}p_j}\,\mathbf{g}_k^{\mathsf{T}}\mathbf{H}\mathbf{f}_j s_j$, and the filtered noise $\mathbf{g}_k^{\mathsf{T}}\mathbf{n}$. Thus, the \ac{SINR} of the $k$-th stream is
\begin{equation}
\gamma_k=\frac{P_\mathrm{T}p_k\big|[\mathbf{E}]_{k,k}\big|^2}{P_\mathrm{T}\sum_{j\ne k}p_j\big|[\mathbf{E}]_{k,j}\big|^2+\varpi^2\big\|[\mathbf{G}]_{k,:}\big\|_2^2},
\label{eq:sinr}
\end{equation}
where $\mathbf{E}\triangleq\mathbf{G}\mathbf{H}\mathbf{F}$ is the effective channel, and the achievable rate is $R=\sum_{k=1}^{N_\mathrm{S}}\log_2(1+\gamma_k)$.

\vspace{-0.3cm}

\subsection{Scattering and Admittance \ac{MiLAC} Representations}
\label{subsec:scattering-rep}

\begin{comment}
\begin{figure} 
         \centering
\includegraphics[width=0.55\linewidth]{MiLAC.pdf}
\vspace{-0.2cm}
       \caption{Example of a $2\times 2$ \ac{MiLAC} network.}
        \label{fig:MiLAC}
        \vspace{-0.6cm}
\end{figure}
\end{comment}

A \ac{MiLAC} can be described either by its scattering matrix, which determines the precoder or the combiner, or by its admittance matrix, which is set by the tunable components and is therefore where the errors occur. Let \(\boldsymbol{\Theta}_\mathrm{T}\) and \(\boldsymbol{\Theta}_\mathrm{R}\) denote the scattering matrices of the transmitter- and receiver-side \acp{MiLAC}, respectively, defined with respect to the reference impedance \(Z_0>0\) of the external ports. Since the \acp{MiLAC} are lossless and reciprocal, their scattering matrices satisfy
\begin{equation}
\label{eq:unitary-sym}
\boldsymbol{\Theta}^\Herm\boldsymbol{\Theta}=\mathbf{I}, \qquad \boldsymbol{\Theta}=\boldsymbol{\Theta}^\T, \end{equation}
for $\boldsymbol{\Theta}\in\{\boldsymbol{\Theta}_\mathrm{T},\boldsymbol{\Theta}_\mathrm{R}\}$. The precoder and the combiner are the off-diagonal blocks $\mathbf{F} = \tfrac{1}{2}[\boldsymbol{\Theta}_\mathrm{T}]_{21}$ and $\mathbf{G} = \tfrac{1}{2}[\boldsymbol{\Theta}_\mathrm{R}]_{12}$~\cite{2025NeriniAnalog1}, so that~\eqref{eq:unitary-sym} also imposes $\|\mathbf{F}\|_2\le\tfrac12$ and $\|\mathbf{G}\|_2\le\tfrac12$.

Let $\Yb\in\{\Yb_\mathrm{T},\Yb_\mathrm{R} \}$ denote the $M\times M$ admittance matrix of a \ac{MiLAC}, with $M\in\{M_\mathrm{T},M_\mathrm{R}\}$. Reciprocity implies $\Yb=\Yb^\T$, and, since a lossless \ac{MiLAC} is purely reactive, $\Yb$ is skew-Hermitian, i.e., purely imaginary. The two descriptions are related by the Cayley transform
\begin{equation}
\boldsymbol{\Theta}=\left(\mathbf{I}+Z_0\Yb\right)^{-1}\left(\mathbf{I}-Z_0\Yb\right)=2\Wb-\mathbf{I},
\label{eq:Y2S}
\end{equation}
where $\Wb\triangleq(\mathbf{I}+Z_0\Yb)^{-1}  \in \{ \Wb_\mathrm{T},\Wb_\mathrm{R} \}$, therefore $\Wb=\tfrac12\left(\mathbf{I}+\boldsymbol{\Theta}\right)$.

\vspace{-0.3cm}
\subsection{Capacity-Achieving Configuration}
\label{subsec:optimal}
We briefly present the capacity-achieving \ac{MiLAC} configuration derived in~\cite{2025neriniCapacity}, which is obtained by considering the \ac{SVD} of the channel, $\mathbf{H}=\mathbf{U}\boldsymbol{\Sigma}\mathbf{V}^\Herm$, with singular values
$\sigma_1\ge\sigma_2\ge\cdots\ge0$ and channel gains
$\lambda_i\triangleq\sigma_i^2$. Defining the matrices
$\Vb\in\mathbb{C}^{N_\mathrm{T}\times N_\mathrm{S}}$ and
$\Ub\in\mathbb{C}^{N_\mathrm{R}\times N_\mathrm{S}}$ containing the first $N_\mathrm{S}$ columns of $\mathbf{V}$ and $\mathbf{U}$,
respectively, and $\Sb\triangleq\diag\left(\sigma_1,\dots,\sigma_{N_\mathrm{S}}\right)$, the optimal analog precoder, combiner, and power allocation of~\cite{2025neriniCapacity} are
\begin{subequations}
\label{eq:FGopt}
\begin{align}
\mathbf{F}^{\mathrm{opt}} &= \tfrac{1}{2}\Vb, \qquad
\mathbf{G}^{\mathrm{opt}} = \tfrac{1}{2}\Ub^\Herm, \\
p_i^{\mathrm{opt}} &= \left(\mu-\frac{4\varpi^2}{P_\mathrm{T}\lambda_i}\right)^+,
\label{eq:wf-powers}
\end{align}
\end{subequations}
where $\mu$ is a constant satisfying $\sum_{i=1}^{N_\mathrm{S}}p_i^{\mathrm{opt}}=1$. 

As a result, the nominal per-stream \ac{SNR} is
\begin{equation}
\gamma_i^{\mathrm{opt}}\triangleq\frac{P_\mathrm{T}p_i^{\mathrm{opt}}\lambda_i}{4\varpi^2},
\label{eq:gamma-star}
\end{equation}
and the achievable rate is $R_{\mathrm{ideal}}=\sum_{i=1}^{N_\mathrm{S}}\log_2\left(1+\gamma_i^{\mathrm{opt}}\right)$. The corresponding effective channel is
\begin{equation}
\mathbf{E}^{\mathrm{opt}}=\mathbf{G}^{\mathrm{opt}} \mathbf{H}\mathbf{F}^{\mathrm{opt}}
=\tfrac14  \diag \left(\sqrt{\lambda_1},\dots,\sqrt{\lambda_{N_\mathrm{S}}}\right).
\label{eq:Heffstar}
\end{equation}
Since $\mathbf{E}^{\mathrm{opt}}$ is diagonal, the channel is decomposed into $N_\mathrm{S}$ interference-free parallel subchannels.

\vspace{-0.3cm}
\subsection{Completion Matrices}
\label{subsec:completion}

The configuration in~\eqref{eq:FGopt} specifies only the precoder and combiner, i.e., the off-diagonal blocks of the scattering matrices. The scattering matrices of the \acp{MiLAC} realizing a given $\mathbf{F}$ and $\mathbf{G}$ therefore take the form~\cite{2025neriniCapacity}
\begin{equation}
\label{eq:Theta-FG}
\boldsymbol{\Theta}_\mathrm{T}=\begin{bmatrix}\boldsymbol{\Xi}_\mathrm{T} & 2\mathbf{F}^\T\\[2pt] 2\mathbf{F} & \boldsymbol{\Psi}_\mathrm{T}\end{bmatrix}, \qquad \boldsymbol{\Theta}_\mathrm{R}=\begin{bmatrix}\boldsymbol{\Xi}_\mathrm{R} & 2\mathbf{G}\\[2pt] 2\mathbf{G}^\T & \boldsymbol{\Psi}_\mathrm{R}\end{bmatrix},
\end{equation}
where the completion matrices $\boldsymbol{\Xi}_\mathrm{T},\boldsymbol{\Xi}_\mathrm{R}\in\mathbb{C}^{N_\mathrm{S}\times N_\mathrm{S}}$, $\boldsymbol{\Psi}_\mathrm{T}\in\mathbb{C}^{N_\mathrm{T}\times N_\mathrm{T}}$, and $\boldsymbol{\Psi}_\mathrm{R}\in\mathbb{C}^{N_\mathrm{R}\times N_\mathrm{R}}$ describe the reflections seen at the \ac{RF}-chain ports and at the antenna ports, and must be chosen such that~\eqref{eq:unitary-sym} holds. Since the rate in~\eqref{eq:sinr} depends only on $\mathbf{F}$ and $\mathbf{G}$, all valid completions achieve the same rate. The completion matrices, however, determine the complete matrix $\Wb=\tfrac12(\mathbf{I}+\boldsymbol{\Theta})$ in~\eqref{eq:Y2S}, and thus the responses of the networks at all their ports. To describe these responses, we partition $\Wb_\mathrm{T}$ and $\Wb_\mathrm{R}$ into their block columns as
\begin{equation}
\label{eq:W-blocks}
\mathbf{W}_\mathrm{T} = \tfrac12 \begin{bmatrix} \Phib_\mathrm{T} & \boldsymbol{\Pi}_\mathrm{T} \end{bmatrix}, \qquad
\mathbf{W}_\mathrm{R} = \tfrac12 \begin{bmatrix} \Phib_\mathrm{R}^* & \boldsymbol{\Pi}_\mathrm{R} \end{bmatrix},
\end{equation}
where the signal-port maps $\Phib_\mathrm{T}\in\mathbb{C}^{M_\mathrm{T}\times N_\mathrm{S}}$ and $\Phib_\mathrm{R}\in\mathbb{C}^{M_\mathrm{R}\times N_\mathrm{S}}$ and the array-port maps $\boldsymbol{\Pi}_\mathrm{T}\in\mathbb{C}^{M_\mathrm{T}\times N_\mathrm{T}}$ and $\boldsymbol{\Pi}_\mathrm{R}\in\mathbb{C}^{M_\mathrm{R}\times N_\mathrm{R}}$, which describe the responses of the networks to excitations at the \ac{RF}-chain ports and at the antenna ports, are
\begin{subequations}
\label{eq:Phi-Pi-gen}
\begin{align}
\Phib_\mathrm{T} & \triangleq\begin{bmatrix}\mathbf{I}_{N_\mathrm{S}}+\boldsymbol{\Xi}_\mathrm{T}\\ 2\mathbf{F}\end{bmatrix}, \qquad \boldsymbol{\Pi}_\mathrm{T} \triangleq\begin{bmatrix}2\mathbf{F}^\T\\ \mathbf{I}_{N_\mathrm{T}}+\boldsymbol{\Psi}_\mathrm{T}\end{bmatrix},\\
\Phib_\mathrm{R}&\triangleq\begin{bmatrix}\mathbf{I}_{N_\mathrm{S}}+\boldsymbol{\Xi}_\mathrm{R}^*\\ 2\mathbf{G}^\Herm\end{bmatrix}, \qquad \boldsymbol{\Pi}_\mathrm{R}\triangleq\begin{bmatrix}2\mathbf{G}\\ \mathbf{I}_{N_\mathrm{R}}+\boldsymbol{\Psi}_\mathrm{R}\end{bmatrix}.
\end{align}
\end{subequations}

For the capacity-achieving configuration in~\eqref{eq:FGopt}, i.e., $2\mathbf{F}=\Vb$ and $2\mathbf{G}=\Ub^\Herm$, the conditions in~\eqref{eq:unitary-sym} restrict the completion matrices as follows:
\begin{subequations}
\label{eq:theta11}
\begin{align}
&\boldsymbol{\Xi}_\mathrm{T}=\boldsymbol{\Xi}_\mathrm{R}=\mathbf{0}_{N_\mathrm{S}\times N_\mathrm{S}},\\
&\boldsymbol{\Psi}_\mathrm{T}=\boldsymbol{\Psi}_\mathrm{T}^\T,\quad \Vb^\Herm\boldsymbol{\Psi}_\mathrm{T}=\mathbf{0}, \quad \boldsymbol{\Psi}_\mathrm{T}\boldsymbol{\Psi}_\mathrm{T}^\Herm=\mathbf{I}-\Vb\Vb^\Herm,\\
&\boldsymbol{\Psi}_\mathrm{R}=\boldsymbol{\Psi}_\mathrm{R}^\T,\quad \boldsymbol{\Psi}_\mathrm{R}\Ub=\mathbf{0},\ \quad  \boldsymbol{\Psi}_\mathrm{R}^\Herm\boldsymbol{\Psi}_\mathrm{R}=\mathbf{I}-\Ub\Ub^\Herm,
\end{align}
\end{subequations}
and the maps in~\eqref{eq:Phi-Pi-gen} reduce to $\Phib_\mathrm{T}=[\mathbf{I}_{N_\mathrm{S}},\Vb^\T]^\T$, $\boldsymbol{\Pi}_\mathrm{T}=[\Vb,\mathbf{I}_{N_\mathrm{T}}+\boldsymbol{\Psi}_\mathrm{T}]^\T$, $\Phib_\mathrm{R}=[\mathbf{I}_{N_\mathrm{S}},\Ub^\T]^\T$, and $\boldsymbol{\Pi}_\mathrm{R}=[\Ub^*,\mathbf{I}_{N_\mathrm{R}}+\boldsymbol{\Psi}_\mathrm{R}]^\T$. In the following section, we analyze $\mathbf{F}^\mathrm{opt}$, $\mathbf{G}^\mathrm{opt}$, and $\mathbf{P}^\mathrm{opt} \triangleq \diag(p_1^\mathrm{opt},\dots,p_{N_{\mathrm{S}}}^\mathrm{opt})$ as the nominal operating point, with a generalized completion satisfying~\eqref{eq:theta11}.

\vspace{-0.3cm}
\section{Rate Analysis Under Intrinsic Impedance Mismatch}
\label{sec:int}

In this section, we analyze the expected rate of the capacity-achieving design under intrinsic impedance mismatch. We model the admittance errors, characterize the perturbations in the precoder and combiner, and derive a closed-form approximation of the expected rate. We then discuss its key properties.

\vspace{-0.3cm}
\subsection{Error Model}
\label{subsec:intrinsic-model}

Tunable reactive components, such as varactor diodes, are specified by manufacturers through guaranteed minimum and maximum capacitance values at given bias voltages, i.e., through the maximum deviation from the nominal value rather than through a probability distribution~\cite{SkyworksSMV1405}. Since the susceptance $\omega C$ of a capacitance $C$ at angular frequency $\omega$ is linear in $C$, a bounded capacitance tolerance yields a bounded susceptance error, and these errors are random and independent across components. When only such bounds are known, the standard approach is to model the quantity as uniformly distributed over the interval they define~\cite[Sec.~4.3.7]{2008JCGMEvaluation}, as is also common for hardware imperfections in wireless systems~\cite{2021QianBeamforming}, since this distribution assigns the same probability to every value within the bounds and, among all distributions supported on a bounded interval, has the maximum entropy, i.e., it introduces no information beyond the bounds.

A lossless \ac{MiLAC} is purely reactive, so that its admittance matrix can be written as $\Yb=\mathrm{j}\mathbf{B}$, where $\mathbf{B}$ is a real and symmetric susceptance matrix. In practice, the realized admittance matrix is $\Yb_\mathrm{act} = \Yb+\Delta\Yb$, where $\Delta\Yb \in \{ \Delta\Yb_\mathrm{T},\Delta\Yb_\mathrm{R} \}$ is the admittance error matrix. Since the component tolerances affect only the reactive part of the components and preserve reciprocity, $\Delta\Yb=\mathrm{j}\Delta\mathbf{B}$, where $\Delta\mathbf{B}=\Delta\mathbf{B}^\T\in\{\Delta\mathbf{B}_\mathrm{T},\Delta\mathbf{B}_\mathrm{R}\}$ is a real matrix of susceptance errors, so that the perturbed networks remain lossless and reciprocal.

We model the free entries of $\Delta\mathbf{B}$ as \ac{i.i.d.} and uniformly distributed over a tolerance interval of half-width $\delta>0$, i.e., $[\Delta\mathbf{B}]_{m,n}\sim\mathcal{U}[-\delta,\delta]$ for $m\le n$. Each entry of $\Delta\Yb$ thus has zero mean and variance $\varsigma^2\triangleq\Ex[|[\Delta\Yb]_{m,n}|^2]=\Ex[[\Delta\mathbf{B}]_{m,n}^2]=\delta^2/3$. For a dimensionless description, we normalize the admittance error by the reference admittance $Y_0\triangleq1/Z_0$ and introduce the error matrix $\Delta\bar{\Yb}\triangleq\Delta\Yb/Y_0 \in \{ \Delta\bar{\Yb}_\mathrm{T},\Delta\bar{\Yb}_\mathrm{R} \}$, whose free entries satisfy $\mathrm{Im}\{[\Delta\bar{\Yb}]_{m,n}\}\sim\mathcal{U}[-\bar{\delta},\bar{\delta}]$, where $\bar{\delta}\triangleq\delta Z_0$ is the normalized tolerance, and whose variance is $\bar{\varsigma}^2\triangleq\varsigma^2Z_0^2=\bar{\delta}^2/3.$

\begin{comment}
\begin{remark}
\label{rem:lossless-perturbed}
Since $\Delta\mathbf{B}$ is real and symmetric, the perturbed admittance matrix remains purely imaginary and symmetric. Thus, the network remains lossless and reciprocal, and $\boldsymbol{\Theta}_\mathrm{act}$ satisfies~\eqref{eq:unitary-sym}. The errors therefore modify the realized beamformers but introduce no losses.
\end{remark}
\end{comment}

\vspace{-0.3cm}
\subsection{Perturbations of the Precoder and Combiner}
\label{subsec:intrinsic-expansion}

The errors occur in the admittance matrix, whereas the beamformers are blocks of the scattering matrix. To relate the two, we expand the Cayley transform~\eqref{eq:Y2S} around the nominal admittance matrix. Since $(\mathbf{I}+Z_0\Yb)\Wb=\mathbf{I}$, the realized network satisfies $\mathbf{I}+Z_0(\Yb+\Delta\Yb)=(\mathbf{I}+Z_0\Yb)(\mathbf{I}+\boldsymbol{\Lambda})$, where $\boldsymbol{\Lambda}\triangleq Z_0\Wb\Delta\Yb \in \{ \boldsymbol{\Lambda}_\mathrm{T}, \boldsymbol{\Lambda}_\mathrm{R} \}$, so that $\Wb_\mathrm{act}\triangleq[\mathbf{I}+Z_0(\Yb+\Delta\Yb)]^{-1}=(\mathbf{I}+\boldsymbol{\Lambda})^{-1}\Wb$. For sufficiently small errors, the Neumann series $(\mathbf{I}+\boldsymbol{\Lambda})^{-1}=\sum_{m=0}^{\infty}(-\boldsymbol{\Lambda})^m$ converges, and the perturbation $\Delta\boldsymbol{\Theta}  \in \{ \Delta\boldsymbol{\Theta}_\mathrm{T}, \Delta\boldsymbol{\Theta}_\mathrm{R} \}$ of the scattering matrix becomes
\begin{equation}
\begin{aligned}
\Delta\boldsymbol{\Theta}&\triangleq\boldsymbol{\Theta}_{\mathrm{act}}-\boldsymbol{\Theta}=2\sum_{m=1}^{\infty}(-Z_0)^m\left(\Wb\Delta\Yb\right)^m\Wb\\
&\approx-2Z_0\Wb\Delta\Yb\Wb,
\end{aligned}
\label{eq:int-first-order}
\end{equation}
where the approximation keeps only the dominant first-order term.
\begin{comment}
Since the nominal network is lossless, $\boldsymbol{\Theta}$ is unitary, and the following relationship satisfies:
\begin{equation}
\|\Wb\|_2 =
\frac{1}{2}
\left\|
\mathbf{I}+\boldsymbol{\Theta}
\right\|_2 \leq
\frac{1}{2}
\left(
\|\mathbf{I}\|_2+
\|\boldsymbol{\Theta}\|_2
\right) =1.
\end{equation}
This also leads to
\begin{equation}
\begin{aligned}
\|\boldsymbol{\Lambda}\|_2
&=
Z_0\|\Wb\Delta\Yb\|_2 \leq
Z_0\|\Wb\|_2\|\Delta\Yb\|_2 \\
& \leq
Z_0\|\Delta\Yb\|_2 =
\frac{\|\Delta\Yb\|_2}{Y_0}.
\end{aligned}
\label{eq:varrho}
\end{equation}
Thus, $\|\Delta\Yb\|_2<Y_0$ is a sufficient condition for $\|\boldsymbol{\Lambda}\|_2<1$. Since every entry of $\Delta\Yb$ is bounded in magnitude by $\delta$, we have
\begin{equation}
\|\Delta\Yb\|_2
\leq
\|\Delta\Yb\|_\mathrm{F}
\leq
\sqrt{M^2\delta^2}
=
M\bar{\delta}\,Y_0 .
\label{eq:dY-norm}
\end{equation}
Hence, $M\bar{\delta}<1$, or equivalently $\sqrt3\,M\bar{\varsigma}<1$ by~\eqref{eq:var-unif-norm}, guarantees $\|\boldsymbol{\Lambda}\|_2<1$ for every realization of the error. Owing to the bounded support of the uniform model, the convergence condition therefore holds deterministically, rather than only with high probability, in the small-error regime considered here. 
\end{comment}

Thus, to first order, the admittance error is filtered on both sides by $\Wb$, so that the sensitivity of a \ac{MiLAC} to mismatch depends on the complete matrix $\Wb$, including the completion matrices, and not only on the beamformers that it realizes. Since $\Wb$ is symmetric,~\eqref{eq:W-blocks} gives $[\Wb_\mathrm{T}]_{2,:}=\tfrac12\boldsymbol{\Pi}_\mathrm{T}^\T$ and $[\Wb_\mathrm{R}]_{1,:}=\tfrac12\Phib_\mathrm{R}^\Herm$ for any completion matrices. Hence, defining the realized precoder and combiner as $\mathbf F_{\mathrm{act}}\triangleq\mathbf F+\Delta\mathbf F$ and $\mathbf{G}_{\mathrm{act}} \triangleq \mathbf{G} + \Delta\mathbf{G}$,~\eqref{eq:int-first-order} gives
\begin{subequations}
\label{eq:dFdG-int}
\begin{align}
\Delta\mathbf{F}&= \tfrac{1}{2} \left[ \Delta\boldsymbol{\Theta}_{\mathrm T} \right]_{21} \approx -\frac{Z_0}{4} \boldsymbol{\Pi}_{\mathrm T}^{\T} \Delta\Yb_{\mathrm T} \Phib_{\mathrm T}, \label{eq:dF-int}\\
\Delta\mathbf{G}&= \tfrac{1}{2} \left[ \Delta\boldsymbol{\Theta}_{\mathrm{R}} \right]_{12} \approx -\frac{Z_0}{4} \Phib_{\mathrm{R}}^{\Herm} \Delta\Yb_{\mathrm{R}} \boldsymbol{\Pi}_{\mathrm{R}}. \label{eq:dG-int}
\end{align}
\end{subequations}

At the nominal operating point, the realized effective channel is $\mathbf{E}_{\mathrm{act}} \triangleq \mathbf{G}_{\mathrm{act}}\mathbf{H}\mathbf{F}_{\mathrm{act}} =\mathbf{E}^\mathrm{opt} + \Delta \mathbf{E}$. Retaining only the first-order terms and using $\mathbf{G}^\mathrm{opt}\mathbf{H}=\tfrac12\Sb\Vb^\Herm$, $\mathbf{H}\mathbf{F}^\mathrm{opt}=\tfrac12\Ub\Sb$, and the identities below~\eqref{eq:theta11}, the perturbation of the effective channel can be expressed as
\begin{align}
\Delta \mathbf{E} &\approx \mathbf{G}^\mathrm{opt}\mathbf{H}\Delta\mathbf{F} + \Delta\mathbf{G}\mathbf{H}\mathbf{F}^\mathrm{opt} \nonumber\\
&= -\frac{Z_0}{8} \left( \Sb\Phib_\mathrm{T}^\Herm \Delta\Yb_\mathrm{T}\Phib_\mathrm{T} + \Phib_\mathrm{R}^\Herm \Delta\Yb_\mathrm{R}\Phib_\mathrm{R}\Sb \right). \label{eq:E-int}
\end{align}
The first term in~\eqref{eq:E-int} is caused by the transmit errors and is scaled by the gain of the stream on which the leakage lands, whereas the second term is caused by the receive errors and is scaled by the gain of the leaking stream.

\vspace{-0.3cm}
\subsection{Expected Rate Under Intrinsic Impedance Mismatch}
\label{subsec:int-rate}

The expected signal, interference, and noise powers depend on the second-order moments of $\Delta\mathbf{E}$ and $\Delta\mathbf{G}$. Since both~\eqref{eq:E-int} and~\eqref{eq:dG-int} are of the form $\mathbf{A}\Delta\Yb\mathbf{B}$ with deterministic $\mathbf{A}$ and $\mathbf{B}$, these moments follow from the next lemma.

\begin{lemma}
\label{lem:moments-int}
Let $\mathbf{A}\in\mathbb{C}^{K\times M}$ and $\mathbf{B}\in\mathbb{C}^{M\times J}$ be deterministic matrices. Then, $\Ex[\mathbf{A}\Delta\Yb\mathbf{B}]=\mathbf{0}$, and the second moments of the entries and the rows of $\mathbf{A}\Delta\Yb\mathbf{B}$ are
\begin{subequations}
\label{eq:closed-moments}
\begin{align}
\Ex\Big[\big|[{\mathbf{A}}\Delta\Yb{\mathbf{B}}]_{k,i}\big|^2\Big]&=\varsigma^2\big[\boldsymbol{\Omega}({\mathbf{A}},{\mathbf{B}})\big]_{k,i},\label{eq:closed-moments-entry}\\
\Ex\Big[\big\|[{\mathbf{A}}\Delta\Yb{\mathbf{B}}]_{k,:}\big\|_2^2\Big]&=\varsigma^2\sum_{i=1}^{J}\big[\boldsymbol{\Omega}({\mathbf{A}},{\mathbf{B}})\big]_{k,i},\label{eq:closed-moments-row}
\end{align}
\end{subequations}
where $\boldsymbol{\Omega}(\mathbf{A},\mathbf{B})\in\mathbb{R}^{K\times J}$ is the matrix with entries $\big[\boldsymbol{\Omega}(\mathbf{A},\mathbf{B})\big]_{k,i}= \big\|[\mathbf{A}]_{k,:}\big\|_2^2\big\|[\mathbf{B}]_{:,i}\big\|_2^2+\big|[\mathbf{A}\mathbf{B}^*]_{k,i}\big|^2 -\sum_{m=1}^{M}\big|[\mathbf{A}]_{k,m}\big|^2\big|[\mathbf{B}]_{m,i}\big|^2.$ Each entry $[\mathbf{A}\Delta\Yb\mathbf{B}]_{k,i}$ is also approximately complex Gaussian by the central limit theorem.
\end{lemma}

\begin{IEEEproof}
See Appendix~\ref{app:admittance-moments}.
\end{IEEEproof}

The first term of $\boldsymbol{\Omega}(\mathbf{A},\mathbf{B})$ is the error power that independent entries of $\Delta\Yb$ would produce, whereas the remaining two terms account for the symmetry of $\Delta\Yb$, which couples its entries $(m,n)$ and $(n,m)$. Applying Lemma~\ref{lem:moments-int} to~\eqref{eq:E-int}, and noting that the transmit and receive errors are independent and zero mean, the expected squared magnitude of the $(k,i)$-th entry of $\Delta\mathbf{E}$ is then
\begin{equation}
\nu_{k,i}^{2} \triangleq \Ex\left[\big|[\Delta\mathbf{E}]_{k,i}\big|^2\right]
= \frac{\bar{\varsigma}^2}{64} \left[ \lambda_k\zeta_{k,i}^{\mathrm{T}} + \lambda_i\zeta_{k,i}^{\mathrm{R}} \right],
\label{eq:nu2-int}
\end{equation}
where $\zeta_{k,i}^{\mathrm{T}}\triangleq[\boldsymbol{\Omega}(\Phib_\mathrm{T}^\Herm,\Phib_\mathrm{T})]_{k,i}$ and $\zeta_{k,i}^{\mathrm{R}}\triangleq[\boldsymbol{\Omega}(\Phib_\mathrm{R}^\Herm,\Phib_\mathrm{R})]_{k,i}$ are the admittance-overlap coefficients of the transmit and receive \acp{MiLAC}, which quantify how strongly the errors couple stream $i$ into stream $k$. Evaluating them with the maps below~\eqref{eq:theta11} gives, for $k\ne i$,
\begin{equation}
\zeta_{k,i}^{\mathrm{T}}=4+\big|\mathbf{v}_k^\T\mathbf{v}_i\big|^2-\sum_{n=1}^{N_\mathrm{T}}\big|[\Vb]_{n,k}\big|^2\big|[\Vb]_{n,i}\big|^2,
\label{eq:zeta}
\end{equation}
where $\mathbf{v}_k$ is the $k$-th column of $\Vb$, and $\zeta_{k,i}^{\mathrm{R}}$ follows by replacing $\Vb$ with $\Ub$. 

\begin{remark}
\label{rem:zeta}
For $k\ne i$, the admittance-overlap coefficients satisfy $7/2\le\zeta_{k,i}^{\mathrm{T}},\zeta_{k,i}^{\mathrm{R}}<5$. The upper bound follows from~\eqref{eq:zeta}, since $|\mathbf{v}_k^\T\mathbf{v}_i|\le\|\mathbf{v}_k\|_2\|\mathbf{v}_i\|_2=1$ by the Cauchy-Schwarz inequality, with equality only if $\mathbf{v}_i$ is proportional to $\mathbf{v}_k^*$, in which case the last term in~\eqref{eq:zeta} is positive. For the lower bound, let $z_n\triangleq[\Vb]_{n,k}^*[\Vb]_{n,i}$, so that the last term in~\eqref{eq:zeta} is $\sum_n|z_n|^2$. Since $\mathbf{v}_k$ and $\mathbf{v}_i$ are orthogonal, $\sum_nz_n=0$, so that no $|z_n|$ exceeds the sum of the others, i.e., $\max_n|z_n|\le\frac12\sum_n|z_n|\le\frac12$, where the last step follows from the Cauchy-Schwarz inequality. Hence, $\sum_n|z_n|^2\le\max_n|z_n|\sum_n|z_n|\le\frac12$, and the same arguments apply to $\zeta_{k,i}^{\mathrm{R}}$ with $\Ub$. These bounds hold for any channel, so that the errors couple every pair of streams with a comparable strength. Moreover, when the singular vectors are spread across the array, i.e., $|[\Vb]_{n,k}|^2\approx1/N_\mathrm{T}$, as for rich-scattering channels and for the steering vector of a \ac{LoS} component, the last two terms in~\eqref{eq:zeta} are of order $1/N_\mathrm{T}$, so that $\zeta_{k,i}^{\mathrm{T}}\approx\bar{\zeta}\triangleq4$, and similarly $\zeta_{k,i}^{\mathrm{R}}\approx\bar{\zeta}$.
\end{remark}

We now evaluate the three terms of the \ac{SINR} in~\eqref{eq:sinr}. For the desired signal, \eqref{eq:Heffstar} gives $[\mathbf{E}_\mathrm{act}]_{k,k}=a_k+[\Delta\mathbf{E}]_{k,k}$, where $a_k\triangleq\sqrt{\lambda_k}/4$, and, since $\Ex[\Delta\mathbf{E}]=\mathbf{0}$ by Lemma~\ref{lem:moments-int}, $\Ex\big[|[\mathbf{E}_\mathrm{act}]_{k,k}|^2\big]=a_k^2+\nu_{k,k}^{2}$. Normalizing the additional power by the nominal noise power $\varpi^2/4$ gives the diagonal perturbation factor
\begin{equation}
\begin{split}
\alpha_k & \triangleq \frac{4P_\mathrm{T}}{\varpi^2} p_k^\mathrm{opt}\nu_{k,k}^{2}
= \frac{\bar{\varsigma}^2}{16} \frac{P_\mathrm{T}}{\varpi^2} p_k^\mathrm{opt}\lambda_k \left( \zeta_{k,k}^{\mathrm{T}} + \zeta_{k,k}^{\mathrm{R}} \right) \\
& = \frac{\bar{\varsigma}^2}{4}\big(\zeta_{k,k}^{\mathrm{T}}+\zeta_{k,k}^{\mathrm{R}}\big)\gamma_k^{\mathrm{opt}}.
\label{eq:alpha-int}
\end{split}
\end{equation}

\begin{comment}
Comparing~\eqref{eq:alpha-int} with~\eqref{eq:gamma-star} shows that
\begin{equation}
\frac{\alpha_k}{\gamma_k^\mathrm{opt}}=\frac{\nu_{k,k}^{2}}{a_k^2}=\frac{\bar{\varsigma}^2}{4}\left(\zeta_{k,k}^{\mathrm{T}}+\zeta_{k,k}^{\mathrm{R}}\right),
\label{eq:alpha-ratio-int}
\end{equation}
which does not depend on the transmit power.
\end{comment}

For the interference, the nominal entry $[\mathbf{E}^\mathrm{opt}]_{k,i}$ is zero for $i\ne k$, so that the leakage of stream $i$ into stream $k$ has power $\Ex\big[|[\mathbf{E}_\mathrm{act}]_{k,i}|^2\big]=\nu_{k,i}^{2}$. Normalizing the total leaked power by $\varpi^2/4$ gives the interference factor
\begin{equation}
\begin{split}
\beta_k &\triangleq \frac{4P_\mathrm{T}}{\varpi^2} \sum_{i\ne k}p_i^\mathrm{opt}\nu_{k,i}^{2}
= \frac{\bar{\varsigma}^2}{16} \frac{P_\mathrm{T}}{\varpi^2} \sum_{i\ne k} p_i^\mathrm{opt} \bigg[ \lambda_k\zeta_{k,i}^{\mathrm{T}}  \\
+ &  \lambda_i\zeta_{k,i}^{\mathrm{R}} \bigg] = \frac{\bar{\varsigma}^2}{4}\sum_{i\neq k}\left(\frac{p_i^{\mathrm{opt}}}{p_k^{\mathrm{opt}}}\,\zeta_{k,i}^{\mathrm{T}}\gamma_k^{\mathrm{opt}}+\zeta_{k,i}^{\mathrm{R}}\gamma_i^{\mathrm{opt}}\right).
\end{split}
\label{eq:beta-int}
\end{equation}
Unlike $\alpha_k$, $\beta_k$ appears in the denominator of the \ac{SINR} and grows with the transmit power, which makes it the main cause of the degradation at high power.

\begin{comment}

The two factors differ only in the power weighting, $p_k^\mathrm{opt}$ against $p_i^\mathrm{opt}$ summed over
$i\ne k$, since both originate from the same perturbed effective channel, one landing on the
stream's own symbol and the other on the remaining streams. The former, however, enters the numerator of the SINR, whereas the latter enters its denominator.
\end{comment}

Finally, the receive errors also change the norms of the rows of the combiner, and thus the noise power. Since $\|[\mathbf{G}^\mathrm{opt}]_{k,:}\|_2^2=\tfrac14$ and $\Ex[\Delta\mathbf{G}]=\mathbf{0}$, the cross term in the expansion of $\|[\mathbf{G}^\mathrm{opt}+\Delta\mathbf{G}]_{k,:}\|_2^2$ vanishes in expectation, and the expected filtered-noise power is $\Ex[\|[\mathbf{G}_{\mathrm{act}}]_{k,:}\|_2^2]=\frac14(1+\chi_k)$, where, by~\eqref{eq:dG-int} and~\eqref{eq:closed-moments-row}, the filtered-noise factor is
\begin{align}
\chi_k &\triangleq 4\,\Ex\Big[\big\|[\Delta\mathbf{G}]_{k,:}\big\|_2^2\Big]=\frac{\bar{\varsigma}^2}{4}\sum_{n=1}^{N_\mathrm{R}}\big[\boldsymbol{\Omega}(\Phib_\mathrm{R}^\Herm,\boldsymbol{\Pi}_\mathrm{R})\big]_{k,n}.
\label{eq:chi-int}
\end{align}
The sum in~\eqref{eq:chi-int} equals $\|[\Phib_\mathrm{R}]_{:,k}\|_2^2\|\boldsymbol{\Pi}_\mathrm{R}\|_\mathrm{F}^2+\|[\Phib_\mathrm{R}]_{:,k}^\T\boldsymbol{\Pi}_\mathrm{R}\|_2^2-\sum_m|[\Phib_\mathrm{R}]_{m,k}|^2\|[\boldsymbol{\Pi}_\mathrm{R}]_{m,:}\|_2^2$. Using $\boldsymbol{\Psi}_\mathrm{R}\Ub=\mathbf{0}$ and the symmetry of $\boldsymbol{\Psi}_\mathrm{R}$, which gives $[\Ub]_{:,k}^\T\boldsymbol{\Psi}_\mathrm{R}=\mathbf{0}$, we have $\|[\Phib_\mathrm{R}]_{:,k}\|_2^2=2$, $\|\boldsymbol{\Pi}_\mathrm{R}\|_\mathrm{F}^2=2N_\mathrm{R}+2\,\mathrm{Re}\{\tr\{\boldsymbol{\Psi}_\mathrm{R}\}\}$, and $\|[\Phib_\mathrm{R}]_{:,k}^\T\boldsymbol{\Pi}_\mathrm{R}\|_2^2=2+2\,\mathrm{Re}\{\sum_n[\Ub]_{n,k}^2\}$, so that~\eqref{eq:chi-int} reduces to
\begin{equation}
\begin{split}
\chi_k &= \bar{\varsigma}^2\Big(N_\mathrm{R}+\theta_k+\mathrm{Re}\{\tr \{\boldsymbol{\Psi}_\mathrm{R} \}\} \\
& -\frac12\sum_n\big|[\Ub]_{n,k}\big|^2\,\mathrm{Re}\big\{[\boldsymbol{\Psi}_\mathrm{R}]_{n,n}\big\}\Big),
\label{eq:chi-int-closed}
\end{split}
\end{equation}
where $\theta_k\triangleq\frac{1}{2}\,\mathrm{Re}\big\{\sum_{n=1}^{N_{\mathrm{R}}}[\overline{\mathbf{U}}]_{n,k}^2\big\}+\frac{1}{4}\sum_{n=1}^{N_{\mathrm{R}}}\big|[\overline{\mathbf{U}}]_{n,k}\big|^2\big[\overline{\mathbf{U}}\overline{\mathbf{U}}^{\mathsf{H}}\big]_{n,n}-\frac{1}{4}$. Unlike $\alpha_k$ and $\beta_k$, $\chi_k$ depends on the completion matrix $\boldsymbol{\Psi}_\mathrm{R}$, which is not determined by the beamformers. To characterize its typical effect, we introduce the following lemma.

\begin{lemma}
\label{lem:completion-average}
Let $\mathbf{M}\in\mathbb{C}^{N_\mathrm{R}\times(N_\mathrm{R}-N_\mathrm{S})}$ be any orthonormal
basis of $\ker(\Ub^\Herm)$. The completions admitted by~\eqref{eq:theta11} can be written as $\boldsymbol{\Psi}_\mathrm{R}=\mathbf{M}^*\mathbf{Q}\mathbf{M}^\Herm,$ with $\mathbf{Q}$ an arbitrary symmetric unitary matrix of order $N_\mathrm{R}-N_\mathrm{S}$.
If $\mathbf{Q}$, or equivalently the basis $\mathbf{M}$, is drawn uniformly at random, then $\Ex\!\left[\boldsymbol{\Psi}_\mathrm{R}\right]=\mathbf{0}.$
\end{lemma}

\begin{IEEEproof}
See Appendix~\ref{app:completion-average}.
\end{IEEEproof}

Using Lemma~\ref{lem:completion-average}, averaging~\eqref{eq:chi-int-closed} over the completions gives the completion-independent filtered-noise factor
\begin{equation}
\Ex_{\boldsymbol{\Psi}_\mathrm{R}}\!\left[\chi_k\right] = \bar{\varsigma}^2\big(N_{\mathrm{R}}+\theta_k\big),
\label{eq:chi-int-avg}
\end{equation}
which replaces $\chi_k$ in the following. Since $|\theta_k|$ is of order one, $\chi_k\approx\bar{\varsigma}^2N_\mathrm{R}$, i.e., the errors make each row of the combiner collect noise from all $N_\mathrm{R}$ antenna ports.

\begin{comment}
This expression can be simplified further when the dominant singular vectors are \emph{spread}, in the sense that their energy is distributed across the array rather than concentrated on a few antennas, so that $|[\Ub]_{n,k}|^2\approx1/N_\mathrm{R}$ for every $n$. This
holds for rich-scattering channels, for which the entries of $\Ub$ behave as those of a Haar
random unitary matrix, and it gives
\begin{subequations}
\label{eq:spread}
\begin{align}
& \sum_n\big|[\Ub]_{n,k}\big|^2\big[\Ub\Ub^\Herm\big]_{n,n}\approx\frac{N_\mathrm{S}}{N_\mathrm{R}},
\label{eq:spread-a}\\
& \sum_n[\Ub]_{n,k}^2=O\!\left(\frac{1}{\sqrt{N_\mathrm{R}}}\right),
\label{eq:spread-b}
\end{align}
\end{subequations}
the first because $[\Ub\Ub^\Herm]_{n,n}\approx N_\mathrm{S}/N_\mathrm{R}$ and
$\sum_n|[\Ub]_{n,k}|^2=1$, and the second because the summands have magnitude $1/N_\mathrm{R}$ and
random phases, so that $N_\mathrm{R}$ of them accumulate as $\sqrt{N_\mathrm{R}}/N_\mathrm{R}$
rather than coherently. Substituting \eqref{eq:spread} into \eqref{eq:chi-int-closed} and
discarding the vanishing term \eqref{eq:spread-b} gives
\begin{equation}
\chi_k\approx\bar{\varsigma}^2\left(N_\mathrm{R}-\frac14+\frac{N_\mathrm{S}}{4N_\mathrm{R}}\right).
\label{eq:chi-int-approx}
\end{equation}
\end{comment}

Thus, replacing the desired-signal, interference, and filtered-noise powers in~\eqref{eq:sinr} with their expected values, and dividing the numerator and denominator by $\varpi^2/4$, the expected \ac{SINR} of the $k$-th stream becomes\footnote{\ac{SINR} expressions in which the errors appear as additive and scaling terms are widely used to characterize the performance of communication systems under different types of errors~\cite{2014BjornsonMassive}.}
\begin{equation}
\bar\gamma_k = \frac{\gamma_k^\mathrm{opt}+\alpha_k}{1+\beta_k+\chi_k},
\label{eq:gammabar-int}
\end{equation}
and the expected achievable rate under admittance errors is $\Ex\left[R_{\mathrm{mismatch}}\right] \approx \sum_{k=1}^{N_\mathrm{S}} \log_2\left(1+\bar\gamma_k\right)$. Hence, the mismatch reduces the \ac{SINR} through the interference $\beta_k$ and the noise enhancement $\chi_k$, whereas $\alpha_k$, whose ratio to $\gamma_k^\mathrm{opt}$ is of order $\bar{\varsigma}^2$, only slightly increases the desired-signal power. Next, we derive the key properties of~\eqref{eq:gammabar-int}.

\vspace{-0.3cm}
\subsection{Key Properties of the Expected Rate}
\label{sec:insights}

\subsubsection{Component Tolerance Scaling}
\label{subsec:ins-tolerance}

All three mismatch factors $\alpha_k$, $\beta_k$, and $\chi_k$ are proportional to $\bar{\varsigma}^2=\bar{\delta}^2/3$. Hence, for small tolerances, the rate degradation is approximately proportional to $\bar{\delta}^2$, whereas it grows more slowly for larger tolerances, since the rate depends logarithmically on the \ac{SINR}. Halving the tolerance divides $\beta_k+\chi_k$ by four, so that, to leading order, the \ac{SINR} of every stream improves by the factor $(1+\beta_k+\chi_k)/\big(1+(\beta_k+\chi_k)/4\big)$. Since $\beta_k$ grows with $P_{\mathrm{T}}$, this factor approaches $4$, i.e., $6$~dB, at high power, which raises the rate by up to $2$~bits/s/Hz per stream, i.e., $2N_{\mathrm{S}}$~bits/s/Hz in total. 

\subsubsection{Rate Ceiling at High Transmit Power}
\label{subsec:ins-ceiling}

Unlike the error-free rate, the expected rate saturates as $P_{\mathrm{T}}$ increases. For a given power allocation, $\gamma_k^{\mathrm{opt}}$, $\alpha_k$, and $\beta_k$ scale linearly with $P_{\mathrm{T}}$, whereas $\chi_k$ is independent of $P_{\mathrm{T}}$. As $P_{\mathrm{T}}\to\infty$, the water-filling allocation~\eqref{eq:wf-powers} approaches uniform power allocation, i.e., $p_k^{\mathrm{opt}}\to 1/N_{\mathrm{S}}$, and $\beta_k$ dominates $1+\chi_k$. Consequently, the expected \ac{SINR} of the $k$-th stream saturates at
\begin{equation}\label{eq:sinr-limit}
\bar{\gamma}_{\mathrm{P},k}^{\infty}=\frac{\lambda_k\big[4+\bar{\varsigma}^2\big(\zeta_{k,k}^{\mathrm{T}}+\zeta_{k,k}^{\mathrm{R}}\big)\big]}{\bar{\varsigma}^2\sum_{i\neq k}\big(\lambda_k\zeta_{k,i}^{\mathrm{T}}+\lambda_i\zeta_{k,i}^{\mathrm{R}}\big)},
\end{equation}
and the expected sum rate approaches the finite rate ceiling $R_{\mathrm{P}}^{\infty}=\sum_{k=1}^{N_{\mathrm{S}}}\log_2\big(1+\bar{\gamma}_{\mathrm{P},k}^{\infty}\big)$. Since both the numerator and the denominator of~\eqref{eq:sinr-limit} scale with the eigenvalues, the ceiling is determined only by the tolerance, the number of streams, and the normalized eigenvalue profile, whereas $P_\mathrm{T}$, $\varpi^2$, and the path loss only determine how fast it is approached. 

\subsubsection{Rate Degradation versus the Number of Antennas}
\label{subsec:ins-array}

Consider $N_{\mathrm{T}}=N_{\mathrm{R}}=N$ with a fixed loading factor $N_\ell=N_{\mathrm{S}}/N$. Increasing $N$ increases both the number of streams and the array gain, so that $R_{\mathrm{ideal}}$ grows linearly with $N$, but it also increases the number of errors that perturb each beamforming weight. To quantify this trade-off, we assume equal power allocation, which water-filling approaches at high per-stream \acp{SNR}, and $\zeta_{k,i}^{\mathrm{T}}\approx\zeta_{k,i}^{\mathrm{R}}\approx\bar{\zeta}$. Then,~\eqref{eq:beta-int} with $\gamma_i^{\mathrm{opt}}/\gamma_k^{\mathrm{opt}}=\lambda_i/\lambda_k$ and $\chi_k\approx\bar{\varsigma}^2N$ give $\beta_k+\chi_k\approx\bar{\varsigma}^2N\rho_k$, where $\rho_k\triangleq1+\frac{\bar{\zeta}N_\ell}{4}\gamma_k^{\mathrm{opt}}\big(1+\bar{\lambda}_{\mathrm{a}}/\lambda_k\big)$ and $\bar{\lambda}_{\mathrm{a}}=\frac{1}{N_{\mathrm{S}}}\sum_{i=1}^{N_{\mathrm{S}}}\lambda_i$ is the average eigenvalue. Since $\alpha_k\ll\gamma_k^{\mathrm{opt}}$, the sum-rate degradation, $\Delta R\triangleq R_{\mathrm{ideal}}-\sum_{k=1}^{N_{\mathrm{S}}}\log_2(1+\bar{\gamma}_k)$, is approximately
\begin{equation}\label{eq:degr-N}
\Delta R\approx\sum_{k=1}^{N_{\mathrm{S}}}\log_2\frac{\big(1+\gamma_k^{\mathrm{opt}}\big)\big(1+\bar{\varsigma}^2N\rho_k\big)}{1+\gamma_k^{\mathrm{opt}}+\bar{\varsigma}^2N\rho_k}.
\end{equation}
For most channel models of interest, such as i.i.d. and correlated Rayleigh fading and Rician fading, all but a few dominant eigenvalues grow linearly with $N$, so that $\gamma_k^{\mathrm{opt}}$ and $\rho_k$ of all but a few streams do not depend on $N$. Hence, since both $R_{\mathrm{ideal}}$ and $\Delta R$ are sums of $N_{\mathrm{S}}$ per-stream terms, and $N$ and $\bar{\delta}$ enter the terms of~\eqref{eq:degr-N} only through $\bar{\delta}^2N$, for a given $P_{\mathrm{T}}$, $N_\ell$, and channel model, the relative degradation $\Delta R/R_{\mathrm{ideal}}$ depends on $N$ and $\bar{\delta}$ only through $\bar{\delta}^2N$. Consequently, doubling $N$ has the same impact as increasing $\bar{\delta}$ by a factor of $\sqrt{2}$, and keeping the relative degradation fixed requires $\bar{\delta}$ to decrease as $1/\sqrt{N}$. Intuitively, the nominal beamforming weights decrease as $1/\sqrt{N}$, whereas the error of each weight, which accumulates contributions from all ports, does not. Moreover,~\eqref{eq:degr-N} shows how the relative degradation grows with $N$. When $\bar{\varsigma}^2N\rho_k\ll1$ for all $k$, $\Delta R\approx\bar{\varsigma}^2N\log_2e\sum_k\rho_k\gamma_k^{\mathrm{opt}}/(1+\gamma_k^{\mathrm{opt}})$, so that $\Delta R/R_{\mathrm{ideal}}$ grows linearly with $N$. For larger $N$, the growth slows down, since the derivative of each term of~\eqref{eq:degr-N} with respect to $\log_2N$ equals $x_k\gamma_k^{\mathrm{opt}}/\big((1+x_k)(1+\gamma_k^{\mathrm{opt}}+x_k)\big)<1$, where $x_k=\bar{\varsigma}^2N\rho_k$. Hence, each doubling of $N$ increases $\Delta R/R_{\mathrm{ideal}}$ by less than $N_{\mathrm{S}}/R_{\mathrm{ideal}}$, i.e., the inverse of the average error-free rate per stream, and $\Delta R/R_{\mathrm{ideal}}$ approaches one, i.e., nearly the entire error-free rate is lost. In contrast, $P_{\mathrm{T}}$, $N_\ell$, and the normalized eigenvalue profile determine $\gamma_k^{\mathrm{opt}}$ and $\rho_k$, and thus how fast the degradation grows with $N$; e.g., a higher $P_{\mathrm{T}}$ increases $\rho_k$ and accelerates the degradation.

\subsubsection{Transmit-Side versus Receive-Side Mismatch}
\label{subsec:ins-asym}

To quantify the distinct roles of the two \acp{MiLAC} anticipated after~\eqref{eq:E-int}, we normalize the inter-stream interference~\eqref{eq:beta-int} by the nominal \ac{SNR} of the stream, which gives
\begin{equation}\label{eq:beta-norm}
\frac{\beta_k}{\gamma_k^{\mathrm{opt}}}=\frac{\bar{\varsigma}^2}{4}\sum_{i\neq k}\frac{p_i^{\mathrm{opt}}}{p_k^{\mathrm{opt}}}\left(\underbrace{\zeta_{k,i}^{\mathrm{T}}}_{\text{transmit}}+\underbrace{\frac{\lambda_i}{\lambda_k}\,\zeta_{k,i}^{\mathrm{R}}}_{\text{receive}}\right).
\end{equation}
Transmit errors occur before the channel: the symbol of stream $i$ leaks into the precoding direction of stream $k$ and propagates through eigenmode $k$ with the same gain $\lambda_k$ as the desired signal, so that the transmit part does not depend on the eigenvalues. Receive errors occur after the channel: stream $i$ arrives with gain $\lambda_i$ and then leaks into the output of stream $k$, so that the receive part is scaled by $\lambda_i/\lambda_k$. Since all overlap coefficients lie between $7/2$ and $5$, this ratio decides which part dominates. For the strongest stream, $\lambda_i/\lambda_1\le 1$ attenuates the receive part, so that the transmit part dominates. For a weak stream, $\lambda_1/\lambda_k\gg 1$ amplifies the receive leakage from the strongest stream, which then dominates and makes the normalized interference of the weak stream much larger than that of the strongest stream. Water-filling reinforces this effect, since it increases $p_1^{\mathrm{opt}}/p_k^{\mathrm{opt}}$ in~\eqref{eq:beta-norm}.

Errors in the receive \ac{MiLAC} generally have a stronger impact on the sum rate, for two reasons. First, only the receive \ac{MiLAC} generates the filtered-noise factor $\chi_k\approx\bar{\varsigma}^2N_{\mathrm{R}}$, which affects every stream, grows with $N_{\mathrm{R}}$, and is present at any transmit power. Second, the receive \ac{MiLAC} leaks more interference power in total. Let $\beta_k^{\mathrm{T}}$ and $\beta_k^{\mathrm{R}}$ denote the parts of~\eqref{eq:beta-int} that contain $\zeta_{k,i}^{\mathrm{T}}$ and $\zeta_{k,i}^{\mathrm{R}}$, so that $\beta_k=\beta_k^{\mathrm{T}}+\beta_k^{\mathrm{R}}$. With $\zeta_{k,i}^{\mathrm{T}}\approx\zeta_{k,i}^{\mathrm{R}}\approx\bar{\zeta}$, summing over the streams gives
\begin{equation}\label{eq:rx-tx-total}
\sum_{k=1}^{N_{\mathrm{S}}}\big(\beta_k^{\mathrm{R}}-\beta_k^{\mathrm{T}}\big)\approx\frac{\bar{\varsigma}^2\bar{\zeta}}{16}\,\frac{P_{\mathrm{T}}}{\varpi^2}\,N_{\mathrm{S}}\left(\sum_{i=1}^{N_{\mathrm{S}}}p_i^{\mathrm{opt}}\lambda_i-\bar{\lambda}_{\mathrm{a}}\right)\ge 0,
\end{equation}
where the inequality follows from Chebyshev's sum inequality, since water-filling assigns more power to stronger eigenmodes, so that $p_i^{\mathrm{opt}}$ and $\lambda_i$ are similarly ordered and $\sum_ip_i^{\mathrm{opt}}\lambda_i\ge\bar{\lambda}_{\mathrm{a}}\sum_ip_i^{\mathrm{opt}}=\bar{\lambda}_{\mathrm{a}}$, with equality only for equal eigenvalues or equal powers. Intuitively, the receive \ac{MiLAC} leaks the received signals, whose powers $p_i^{\mathrm{opt}}\lambda_i$ are dominated by the strongest streams, and this leakage falls mostly on the weak streams, which lose a large fraction of their rate, whereas the transmit leakage falls mostly on the strong streams, whose high \ac{SINR} makes their rate less sensitive. Hence, if components of different tolerances are available, the tighter ones should be used at the receiver.

\subsubsection{Impact of Additional Streams}
\label{subsec:ins-streams}

The leakage into stream $k$ depends on the power transmitted on the other streams rather than on their number. With $\zeta_{k,i}^{\mathrm{T}}\approx\zeta_{k,i}^{\mathrm{R}}\approx\bar{\zeta}$,~\eqref{eq:beta-int} gives
\begin{equation}\label{eq:beta-NS}
\beta_k\approx\frac{\bar{\varsigma}^2\bar{\zeta}P_{\mathrm{T}}}{16\varpi^2}\bigg[\big(1-p_k^{\mathrm{opt}}\big)\lambda_k+\sum_{i\ne k}p_i^{\mathrm{opt}}\lambda_i\bigg]<\frac{\bar{\varsigma}^2\bar{\zeta}P_{\mathrm{T}}}{16\varpi^2}\big(\lambda_k+\lambda_1\big),
\end{equation}
where the inequality follows from $\sum_ip_i^{\mathrm{opt}}=1$ and $\lambda_i\le\lambda_1$. Hence, adding streams only redistributes a fixed total power, the leakage into each stream remains bounded by a value that does not depend on $N_{\mathrm{S}}$, and the relative degradation depends only moderately on $N_{\mathrm{S}}$. Moreover, water-filling activates only the eigenmodes whose gains exceed the water level, so that increasing $N_{\mathrm{S}}$ beyond the number of active eigenmodes changes neither the error-free nor the expected rate.

The leakage, however, reduces the value of weak streams. Activating a weak stream $i$ adds the term $\frac{\bar{\varsigma}^2}{16}\frac{P_{\mathrm{T}}}{\varpi^2}p_i^{\mathrm{opt}}\lambda_1\zeta_{1,i}^{\mathrm{T}}$ to the leakage into the strongest stream, which grows with $P_{\mathrm{T}}$. In contrast, the weak stream receives the leakage of the strongest stream, $\beta_i\ge\frac{\bar{\varsigma}^2}{4}\zeta_{i,1}^{\mathrm{R}}\gamma_1^{\mathrm{opt}}$, so that $\bar{\gamma}_i\lesssim\gamma_i^{\mathrm{opt}}/\beta_i\le 4p_i^{\mathrm{opt}}\lambda_i/\big(\bar{\varsigma}^2\zeta_{i,1}^{\mathrm{R}}p_1^{\mathrm{opt}}\lambda_1\big)$, which does not grow with $P_{\mathrm{T}}$. Hence, at high power, the rate lost on the strongest stream can exceed that contributed by the weak stream, and assigning less power to weak streams than water-filling can increase the rate. 

% =====================================================================
\vspace{-0.3cm}
\section{Robust Design}
\label{sec:robust}

Section~\ref{sec:int} showed that the capacity-achieving design, which ignores the admittance errors, can suffer a substantial rate degradation. In this section, we account for the errors and optimize the \acp{MiLAC} and the power allocation to maximize the expected sum rate.

Without errors, the rate in~\eqref{eq:sinr} depends only on $\mathbf{F}$ and $\mathbf{G}$, while the completion matrices can be chosen arbitrarily, provided that~\eqref{eq:unitary-sym} is satisfied. The design therefore reduces to a conventional \ac{MIMO} precoding problem with $N_\mathrm{S}$ interference-free parallel subchannels. The admittance errors break this structure in three ways. First, as shown by~\eqref{eq:dF-int} and~\eqref{eq:dG-int}, the perturbations of the beamformers depend on the complete matrices $\Wb_\mathrm{T}$ and $\Wb_\mathrm{R}$, including the completion matrices. Thus, the design can no longer be reduced to $\mathbf{F}$ and $\mathbf{G}$. Second, the errors introduce interference between every pair of streams that cannot be eliminated through the choice of beamformers, as shown by~\eqref{eq:nu2-int} for the capacity-achieving design. Consequently, the channel can no longer be decomposed into parallel subchannels, and the \ac{SINR} of each stream depends on the power allocated to all streams. Water-filling is therefore no longer optimal, and the power allocation must be jointly optimized with the \acp{MiLAC}. Third, the expected error powers are fourth-order functions of the scattering matrices, which are constrained to be symmetric and unitary. Consequently, the robust design does not admit a closed-form solution, and we optimize the complete scattering matrices and the power allocation iteratively.

\vspace{-0.3cm}

\subsection{Problem Formulation}
\label{subsec:robust-rate}

To formulate the problem, we need the expected rate for arbitrary scattering matrices, rather than only at the capacity-achieving configuration as in Section~\ref{sec:int}. The realized effective channel is $\mathbf{E}_\mathrm{act}=\mathbf{G}\mathbf{H}\mathbf{F}+\mathbf{G}\mathbf{H}\Delta\mathbf{F}+\Delta\mathbf{G}\mathbf{H}\mathbf{F}+\Delta\mathbf{G}\mathbf{H}\Delta\mathbf{F}$, where the last term is of second order. Since~\eqref{eq:dF-int} and~\eqref{eq:dG-int} hold for any completion matrices with the maps in~\eqref{eq:Phi-Pi-gen}, the first-order error terms are
\begin{subequations}
\label{eq:error-terms}
\begin{align}
\mathbf{G}\mathbf{H}\Delta\mathbf{F}&\approx-\frac{Z_0}{4}\big(\mathbf{G}\mathbf{H}\boldsymbol{\Pi}_\mathrm{T}^\T\big)\Delta\Yb_\mathrm{T}\Phib_\mathrm{T},\label{eq:err-tx}\\
\Delta\mathbf{G}\mathbf{H}\mathbf{F}&\approx-\frac{Z_0}{4}\Phib_\mathrm{R}^\Herm\Delta\Yb_\mathrm{R}\big(\boldsymbol{\Pi}_\mathrm{R}\mathbf{H}\mathbf{F}\big).\label{eq:err-rx}
\end{align}
\end{subequations}
Applying Lemma~\ref{lem:moments-int} to~\eqref{eq:error-terms} and~\eqref{eq:dG-int}, retaining only the first-order terms, and noting that the transmit and receive errors are independent and zero mean, so that all cross terms vanish, the expected power of the $(k,i)$-th entry of $\mathbf{E}_\mathrm{act}$ and the expected squared norm of the $k$-th row of $\mathbf{G}_\mathrm{act}$ are
\begin{subequations}
\label{eq:qN-final}
\begin{align}
\bar{E}_{k,i}&\triangleq\Ex\Big[\big|[\mathbf{E}_\mathrm{act}]_{k,i}\big|^2\Big]=\big|[\mathbf{G}\mathbf{H}\mathbf{F}]_{k,i}\big|^2+\frac{\bar{\varsigma}^2}{16}\left(\omega_{k,i}^\mathrm{T}+\omega_{k,i}^\mathrm{R}\right),\label{eq:q-final}\\
\bar{G}_k&\triangleq\Ex\Big[\big\|[\mathbf{G}_\mathrm{act}]_{k,:}\big\|_2^2\Big]=\big\|[\mathbf{G}]_{k,:}\big\|_2^2+\frac{\bar{\varsigma}^2}{16}\,\omega_k^\mathrm{N},\label{eq:N-final}
\end{align}
\end{subequations}
where the error-coupling coefficients of the transmit \ac{MiLAC}, of the receive \ac{MiLAC}, and of the filtered noise, which generalize $\nu_{k,i}^2$ and $\chi_k$ to arbitrary scattering matrices, are
\begin{subequations}
\label{eq:omega-final}
\begin{align}
\omega_{k,i}^\mathrm{T}&\triangleq\Big[\boldsymbol{\Omega}\big(\mathbf{G}\mathbf{H}\boldsymbol{\Pi}_\mathrm{T}^\T,\Phib_\mathrm{T}\big)\Big]_{k,i},\label{eq:omega-T}\\
\omega_{k,i}^\mathrm{R}&\triangleq\Big[\boldsymbol{\Omega}\big(\Phib_\mathrm{R}^\Herm,\boldsymbol{\Pi}_\mathrm{R}\mathbf{H}\mathbf{F}\big)\Big]_{k,i},\label{eq:omega-R}\\
\omega_k^\mathrm{N}&\triangleq\sum_{n=1}^{N_\mathrm{R}}\Big[\boldsymbol{\Omega}\big(\Phib_\mathrm{R}^\Herm,\boldsymbol{\Pi}_\mathrm{R}\big)\Big]_{k,n}.\label{eq:omega-N}
\end{align}
\end{subequations}

We then express all these quantities in terms of $\boldsymbol{\Theta}_\mathrm{T}$ and $\boldsymbol{\Theta}_\mathrm{R}$. Let $\mathbf{S}_{\mathrm{T},1}\triangleq[\mathbf{I}_{N_\mathrm{S}},\mathbf{0}]^\T\in\mathbb{R}^{M_\mathrm{T}\times N_\mathrm{S}}$ and $\mathbf{S}_{\mathrm{T},2}\triangleq[\mathbf{0},\mathbf{I}_{N_\mathrm{T}}]^\T\in\mathbb{R}^{M_\mathrm{T}\times N_\mathrm{T}}$ select the \ac{RF}-chain ports and the antenna ports of the transmit \ac{MiLAC}, respectively, and let $\mathbf{S}_{\mathrm{R},1}\in\mathbb{R}^{M_\mathrm{R}\times N_\mathrm{S}}$ and $\mathbf{S}_{\mathrm{R},2}\in\mathbb{R}^{M_\mathrm{R}\times N_\mathrm{R}}$ be their counterparts for the receive \ac{MiLAC}. We also define $\widetilde{\mathbf{H}}\triangleq\mathbf{S}_{\mathrm{R},2}\mathbf{H}\mathbf{S}_{\mathrm{T},2}^\T\in\mathbb{C}^{M_\mathrm{R}\times M_\mathrm{T}}$, i.e., the channel between all the ports of the two \acp{MiLAC}, whose rows and columns associated with the \ac{RF}-chain ports are zero.
Then, since $\mathbf{F}=\tfrac12[\boldsymbol{\Theta}_\mathrm{T}]_{21}$ and $\mathbf{G}=\tfrac12[\boldsymbol{\Theta}_\mathrm{R}]_{12}$, the precoder, the combiner, and the error-free effective channel are
\begin{subequations}
\label{eq:FG-Theta}
\begin{align}
&\mathbf{F}=\tfrac12\mathbf{S}_{\mathrm{T},2}^\T\boldsymbol{\Theta}_\mathrm{T}\mathbf{S}_{\mathrm{T},1}, \qquad
\mathbf{G}=\tfrac12\mathbf{S}_{\mathrm{R},1}^\T\boldsymbol{\Theta}_\mathrm{R}\mathbf{S}_{\mathrm{R},2},\label{eq:G-Theta}\\
&\mathbf{G}\mathbf{H}\mathbf{F}=\tfrac14\mathbf{S}_{\mathrm{R},1}^\T\boldsymbol{\Theta}_\mathrm{R}\widetilde{\mathbf{H}}\boldsymbol{\Theta}_\mathrm{T}\mathbf{S}_{\mathrm{T},1},\label{eq:E-Theta}
\end{align}
\end{subequations}
and, by~\eqref{eq:W-blocks}, the port maps are
\begin{subequations}
\label{eq:maps-Theta}
\begin{align}
\Phib_\mathrm{T}&=2\Wb_\mathrm{T}\mathbf{S}_{\mathrm{T},1}, \qquad
\boldsymbol{\Pi}_\mathrm{T}^\T=2\mathbf{S}_{\mathrm{T},2}^\T\Wb_\mathrm{T},\label{eq:map-PhiT}\\
\Phib_\mathrm{R}^\Herm&=2\mathbf{S}_{\mathrm{R},1}^\T\Wb_\mathrm{R}, \qquad
\boldsymbol{\Pi}_\mathrm{R}=2\Wb_\mathrm{R}\mathbf{S}_{\mathrm{R},2}.\label{eq:map-PiR}
\end{align}
\end{subequations}
Since $\mathbf{F}=\mathbf{S}_{\mathrm{T},2}^\T\Wb_\mathrm{T}\mathbf{S}_{\mathrm{T},1}$ and $\mathbf{G}=\mathbf{S}_{\mathrm{R},1}^\T\Wb_\mathrm{R}\mathbf{S}_{\mathrm{R},2}$ are also selections of $\Wb_\mathrm{T}$ and $\Wb_\mathrm{R}$, all matrices in~\eqref{eq:qN-final} and~\eqref{eq:omega-final} are linear functions of $\Wb_\mathrm{T}=\tfrac12(\mathbf{I}+\boldsymbol{\Theta}_\mathrm{T})$ and $\Wb_\mathrm{R}=\tfrac12(\mathbf{I}+\boldsymbol{\Theta}_\mathrm{R})$, or products of such functions, and we write $\bar{E}_{k,i}(\boldsymbol{\Theta}_\mathrm{T},\boldsymbol{\Theta}_\mathrm{R})$ and $\bar{G}_k(\boldsymbol{\Theta}_\mathrm{R})$. Thus, the expected \ac{SINR} of the $k$-th stream is
\begin{equation}
\bar\gamma_k(\boldsymbol{\Theta}_\mathrm{T},\boldsymbol{\Theta}_\mathrm{R},\mathbf{p})=\frac{P_\mathrm{T}p_k\bar{E}_{k,k}(\boldsymbol{\Theta}_\mathrm{T},\boldsymbol{\Theta}_\mathrm{R})}{P_\mathrm{T}\sum\limits_{i\ne k}p_i\bar{E}_{k,i}(\boldsymbol{\Theta}_\mathrm{T},\boldsymbol{\Theta}_\mathrm{R})+\varpi^2\bar{G}_k(\boldsymbol{\Theta}_\mathrm{R})},
\label{eq:gammabar-gen}
\end{equation}
and the expected rate is $\bar R(\boldsymbol{\Theta}_\mathrm{T},\boldsymbol{\Theta}_\mathrm{R},\mathbf{p})=\sum_{k}\log_2\big(1+\bar\gamma_k(\boldsymbol{\Theta}_\mathrm{T},\boldsymbol{\Theta}_\mathrm{R},\mathbf{p})\big)$.

The robust design problem is then formulated as
\begin{subequations}
\label{eq:robust-problem}
\begin{align}
\max_{\boldsymbol{\Theta}_\mathrm{T},\boldsymbol{\Theta}_\mathrm{R},\mathbf{p}}\;
& \bar R(\boldsymbol{\Theta}_\mathrm{T},\boldsymbol{\Theta}_\mathrm{R},\mathbf{p})\label{eq:rp-obj}\\
\mathrm{s.t.}\;
& \boldsymbol{\Theta}_\mathrm{T}\ \text{and}\ \boldsymbol{\Theta}_\mathrm{R}\ \text{satisfy~\eqref{eq:unitary-sym}},\label{eq:rp-net}\\
& \mathbf{p}\ge\mathbf{0},\label{eq:rp-pos}\\
& \mathbf{1}^\T\mathbf{p}=1.\label{eq:rp-sum}
\end{align}
\end{subequations}
This problem is non-convex, since its objective is not concave,~\eqref{eq:rp-net} defines a non-convex set, and the variables are coupled. We therefore propose a method to find a stationary point of~\eqref{eq:robust-problem}.

\begin{comment}
\begin{remark}
\label{rem:completion-dependence}
In the absence of errors, the rate depends only on $\mathbf{F}$, $\mathbf{G}$, and $\mathbf{p}$. Thus, different scattering matrices that yield the same $\mathbf{F}$ and $\mathbf{G}$ achieve the same rate, even if their reflection blocks $\boldsymbol{\Xi}$ and $\boldsymbol{\Psi}$ differ. However, these reflection blocks correspond to different admittance matrices through~\eqref{eq:Y2S}, which can have different sensitivities to admittance errors. Specifically,~\eqref{eq:dF-int} shows that this sensitivity depends on the port voltages induced by antenna and \ac{RF}-chain excitations, which vary with the reflection blocks. Consequently, networks with the same nominal rate can experience different rate losses under the same admittance errors. Therefore, optimizing the complete scattering matrices provides additional freedom to mitigate the effects of hardware imperfections.
\end{remark}
\end{comment}

\vspace{-0.3cm}
\subsection{Proposed Solution}
\label{subsec:robust-solution}

Two features of~\eqref{eq:robust-problem} make projected gradient ascent a natural choice: its objective is differentiable with closed-form gradients, and, although its feasible set is non-convex, the projection onto it is available in closed form, as shown in Lemma~\ref{lem:projection}. To accelerate convergence, we take each gradient step from an extrapolated point that combines the two previous iterates. Since~\eqref{eq:robust-problem} is non-convex, this momentum alone does not guarantee that the objective increases, and we therefore use the monotone accelerated scheme of~\cite{li2015accelerated}, which, at every iteration, compares the accelerated step with a plain projected gradient step taken from the current iterate, referred to as \emph{the safeguard step}, and keeps the better one.

We collect the three variables into the triple $\mathbf{x}\triangleq(\boldsymbol{\Theta}_\mathrm{T},\boldsymbol{\Theta}_\mathrm{R},\mathbf{p})$. The extrapolated point $\mathbf{y}$, the accelerated candidate $\mathbf{z}$, and the safeguard candidate $\mathbf{v}$ are triples of the same form, whose components associated with the variable $\boldsymbol{\vartheta}\in\{\boldsymbol{\Theta}_\mathrm{T},\boldsymbol{\Theta}_\mathrm{R},\mathbf{p}\}$ are denoted by $\mathbf{y}_{\boldsymbol{\vartheta}}$, $\mathbf{z}_{\boldsymbol{\vartheta}}$, and $\mathbf{v}_{\boldsymbol{\vartheta}}$. At every iteration, all three variables are updated simultaneously, using partial gradients evaluated at the same point, and, since the feasible set of~\eqref{eq:robust-problem} is the Cartesian product of the feasible sets of the three variables, the projection decomposes into one projection per variable. Hence, every iteration applies the following steps to each variable $\boldsymbol{\vartheta}$:

\emph{Step 1: Extrapolation:} The variable is first extrapolated as
\begin{equation}
\mathbf{y}_{\boldsymbol{\vartheta}}^{(t)}=\boldsymbol{\vartheta}^{(t)}
+\frac{\varphi_{t-1}}{\varphi_t}\big(\mathbf{z}_{\boldsymbol{\vartheta}}^{(t)}-\boldsymbol{\vartheta}^{(t)}\big)
+\frac{\varphi_{t-1}-1}{\varphi_t}\big(\boldsymbol{\vartheta}^{(t)}-\boldsymbol{\vartheta}^{(t-1)}\big),
\label{eq:generic-extrapolation}
\end{equation}
where $t$ is the iteration number and $\varphi_t$ is the momentum sequence updated as $\varphi_{t+1}=\frac12(1+\sqrt{1+4\varphi_t^2})$, with $\varphi_0=0$ and $\varphi_1=1$.

 \emph{Step 2: Gradient ascent update:} Two candidates are produced: the accelerated candidate, by a gradient step from the extrapolated point $\mathbf{y}^{(t)}$, and the safeguard candidate, by a gradient step from the current iterate $\mathbf{x}^{(t)}$, which, after the projection in Step~3 and with the step size selected by backtracking, never decreases the objective. They are given by
\begin{subequations}
\label{eq:generic-update}
\begin{align}
\bar{\mathbf{z}}_{\boldsymbol{\vartheta}}^{(t+1)}&=\mathbf{y}_{\boldsymbol{\vartheta}}^{(t)}+\alpha_\mathrm{y}^{(t)}\nabla_{\boldsymbol{\vartheta}}\bar R\big(\mathbf{y}^{(t)}\big),\label{eq:generic-update-acc}\\
\bar{\mathbf{v}}_{\boldsymbol{\vartheta}}^{(t+1)}&=\boldsymbol{\vartheta}^{(t)}+\alpha_\mathrm{x}^{(t)}\nabla_{\boldsymbol{\vartheta}}\bar R\big(\mathbf{x}^{(t)}\big),\label{eq:generic-update-pg}
\end{align}
\end{subequations}
where $\alpha_\mathrm{y}^{(t)}>0$ and $\alpha_\mathrm{x}^{(t)}>0$ are the step sizes, and $\nabla_{\boldsymbol{\vartheta}}\bar R(\mathbf{u})$ is the partial gradient with respect to $\boldsymbol{\vartheta}$ at the point $\mathbf{u}$, i.e., with the other two variables fixed at their values in $\mathbf{u}$.

 \emph{Step 3: Projection:} The updated points are projected onto the feasible set of the variable as
\begin{subequations}
\label{eq:generic-projection}
\begin{align}
\mathbf{z}_{\boldsymbol{\vartheta}}^{(t+1)}&=\mathcal{T}_{\boldsymbol{\vartheta}}\big(\bar{\mathbf{z}}_{\boldsymbol{\vartheta}}^{(t+1)}\big),\label{eq:generic-projection-acc}\\
\mathbf{v}_{\boldsymbol{\vartheta}}^{(t+1)}&=\mathcal{T}_{\boldsymbol{\vartheta}}\big(\bar{\mathbf{v}}_{\boldsymbol{\vartheta}}^{(t+1)}\big),\label{eq:generic-projection-pg}
\end{align}
\end{subequations}
where $\mathcal{T}_{\boldsymbol{\vartheta}}(\cdot)$ is the Euclidean projection onto the feasible set of $\boldsymbol{\vartheta}$. Once the three variables have been updated and projected, the next iterate is the candidate triple achieving the larger expected rate, i.e.,
\begin{equation}
\mathbf{x}^{(t+1)}=
\begin{cases}
\mathbf{z}^{(t+1)}, & \text{if }\bar R\big(\mathbf{z}^{(t+1)}\big)\ge\bar R\big(\mathbf{v}^{(t+1)}\big),\\
\mathbf{v}^{(t+1)}, & \text{otherwise}.
\end{cases}
\label{eq:selection}
\end{equation}
Hence, the method only requires the partial gradient $\nabla_{\boldsymbol{\vartheta}}\bar R$ and the projection $\mathcal{T}_{\boldsymbol{\vartheta}}(\cdot)$ of each variable, which we derive next.

We first introduce the quantities shared by the three gradients. Let $T_k\triangleq P_\mathrm{T}\sum_ip_i\bar{E}_{k,i}+\varpi^2\bar{G}_k$ and $J_k\triangleq T_k-P_\mathrm{T}p_k\bar{E}_{k,k}$ denote the total received power and the interference-plus-noise power of the $k$-th stream, respectively, so that
\begin{equation}
\bar R=\frac{1}{\ln2}\sum_{k=1}^{N_\mathrm{S}}\big(\ln T_k-\ln J_k\big).
\label{eq:Rbar-TJ}
\end{equation}
Differentiating~\eqref{eq:Rbar-TJ} with $\mathrm{d}T_k=P_\mathrm{T}\sum_ip_i\,\mathrm{d}\bar{E}_{k,i}+\varpi^2\mathrm{d}\bar{G}_k$ and $\mathrm{d}J_k=P_\mathrm{T}\sum_{i\ne k}p_i\,\mathrm{d}\bar{E}_{k,i}+\varpi^2\mathrm{d}\bar{G}_k$ gives
\begin{equation}
\mathrm{d}\bar R=\sum_{k=1}^{N_\mathrm{S}}\sum_{i=1}^{N_\mathrm{S}}\upsilon_{k,i}\,\mathrm{d}\bar{E}_{k,i}+\sum_{k=1}^{N_\mathrm{S}}\kappa_k\,\mathrm{d}\bar{G}_k,
\label{eq:dR-weighted}
\end{equation}
with the weights
\begin{equation}
\label{eq:weights}
\upsilon_{k,i}\triangleq\frac{P_\mathrm{T}p_i}{\ln2}\left(\frac{1}{T_k}-\frac{\mathds{1}_{i\ne k}}{J_k}\right), \quad
\kappa_k\triangleq\frac{\varpi^2}{\ln2}\left(\frac{1}{T_k}-\frac{1}{J_k}\right).
\end{equation}
Since $T_k>J_k$, the weight $\upsilon_{k,k}$ of the desired signal is positive, whereas the weights $\upsilon_{k,i}$, $i\ne k$, and $\kappa_k$ of the interference and the noise are negative, so that, locally, maximizing $\bar R$ amounts to maximizing a weighted sum of powers that rewards the desired signals and penalizes the interference and the noise. We collect the weights in $\boldsymbol{\Upsilon}\in\mathbb{R}^{N_\mathrm{S}\times N_\mathrm{S}}$, with $[\boldsymbol{\Upsilon}]_{k,i}=\upsilon_{k,i}$, and in $\boldsymbol{\kappa}\triangleq[\kappa_1,\dots,\kappa_{N_\mathrm{S}}]^\T$, and we scale them by the error variance as $\tilde{\boldsymbol{\Upsilon}}\triangleq\frac{\bar{\varsigma}^2}{16}\boldsymbol{\Upsilon}$ and $\tilde{\boldsymbol{\kappa}}\triangleq\frac{\bar{\varsigma}^2}{16}\boldsymbol{\kappa}\mathbf{1}_{N_\mathrm{R}}^\T\in\mathbb{R}^{N_\mathrm{S}\times N_\mathrm{R}}$.

The error terms $\omega^\mathrm{T}_{k,i}$, $\omega^\mathrm{R}_{k,i}$, and $\omega^\mathrm{N}_k$ in~\eqref{eq:omega-final} are entries of $\boldsymbol{\Omega}(\mathbf{A},\mathbf{B})$ evaluated at the pairs $(\mathbf{C}_\mathrm{T},\Phib_\mathrm{T})$, $(\Phib_\mathrm{R}^\Herm,\mathbf{C}_\mathrm{R})$, and $(\Phib_\mathrm{R}^\Herm,\boldsymbol{\Pi}_\mathrm{R})$, respectively, where $\mathbf{C}_\mathrm{T}\triangleq\mathbf{G}\mathbf{H}\boldsymbol{\Pi}_\mathrm{T}^\T$ and $\mathbf{C}_\mathrm{R}\triangleq\boldsymbol{\Pi}_\mathrm{R}\mathbf{H}\mathbf{F}$. For a real weight matrix $\boldsymbol{\Gamma}$, let $\phi_{\boldsymbol{\Gamma}}(\mathbf{A},\mathbf{B})\triangleq\sum_{k,i}[\boldsymbol{\Gamma}]_{k,i}[\boldsymbol{\Omega}(\mathbf{A},\mathbf{B})]_{k,i}$ denote the corresponding weighted sum, and let $\nabla\phi_{\boldsymbol{\Gamma}}(\mathbf{A},\mathbf{B})\triangleq\big(\partial\phi_{\boldsymbol{\Gamma}}/\partial\mathbf{A}^*,\partial\phi_{\boldsymbol{\Gamma}}/\partial\mathbf{B}^*\big)$ denote its derivatives, which are given in closed form by~\eqref{eq:dphi-A} and~\eqref{eq:dphi-B} in Appendix~\ref{app:gradient}. Evaluating them for the three error terms gives the six matrices
\begin{subequations}
\label{eq:backprop}
\begin{align}
\big(\dot{\mathbf{A}}_\mathrm{T},\dot{\mathbf{B}}_\mathrm{T}\big)&\triangleq\nabla\phi_{\tilde{\boldsymbol{\Upsilon}}}\big(\mathbf{C}_\mathrm{T},\Phib_\mathrm{T}\big),\label{eq:backprop-T}\\
\big(\dot{\mathbf{A}}_\mathrm{R},\dot{\mathbf{B}}_\mathrm{R}\big)&\triangleq\nabla\phi_{\tilde{\boldsymbol{\Upsilon}}}\big(\Phib_\mathrm{R}^\Herm,\mathbf{C}_\mathrm{R}\big),\label{eq:backprop-R}\\
\big(\dot{\mathbf{A}}_\mathrm{N},\dot{\mathbf{B}}_\mathrm{N}\big)&\triangleq\nabla\phi_{\tilde{\boldsymbol{\kappa}}}\big(\Phib_\mathrm{R}^\Herm,\boldsymbol{\Pi}_\mathrm{R}\big).\label{eq:backprop-N}
\end{align}
\end{subequations}
Here, $\mathbf{E}=\mathbf{G}\mathbf{H}\mathbf{F}$, and the gradient with respect to a scattering matrix is taken in the sense that $\mathrm{d}\bar R=\mathrm{Re}\{\tr\{(\nabla_{\boldsymbol{\Theta}}\bar R)^\Herm\mathrm{d}\boldsymbol{\Theta}\}\}$ for every symmetric $\mathrm{d}\boldsymbol{\Theta}$.

\subsubsection{Transmit Scattering Matrix}
The expected rate $\bar R$ depends on $\boldsymbol{\Theta}_\mathrm{T}$ through three paths: (1) the precoder $\mathbf{F}$, which determines the nominal effective channel $\mathbf{E}$ and, through $\mathbf{C}_\mathrm{R}$, the receive error term $\omega^\mathrm{R}_{k,i}$; (2) the matrix $\mathbf{C}_\mathrm{T}$ of the transmit error term $\omega^\mathrm{T}_{k,i}$, through $\boldsymbol{\Pi}_\mathrm{T}$; and (3) the matrix $\Phib_\mathrm{T}$. Collecting path (1) into
\begin{equation}
\dot{\mathbf{F}}\triangleq(\mathbf{G}\mathbf{H})^\Herm\big(\boldsymbol{\Upsilon}\odot\mathbf{E}\big)+(\boldsymbol{\Pi}_\mathrm{R}\mathbf{H})^\Herm\dot{\mathbf{B}}_\mathrm{R},
\label{eq:F-dot}
\end{equation}
and propagating the three paths to $\Wb_\mathrm{T}$ through~\eqref{eq:maps-Theta} gives
\begin{equation}
\mathbf{D}_\mathrm{T}\triangleq\mathbf{S}_{\mathrm{T},2}\dot{\mathbf{F}}\mathbf{S}_{\mathrm{T},1}^\T
+2\mathbf{S}_{\mathrm{T},2}(\mathbf{G}\mathbf{H})^\Herm\dot{\mathbf{A}}_\mathrm{T}
+2\dot{\mathbf{B}}_\mathrm{T}\mathbf{S}_{\mathrm{T},1}^\T.
\label{eq:DT}
\end{equation}
Since $\boldsymbol{\Theta}_\mathrm{T}=2\Wb_\mathrm{T}-\mathbf{I}$ and the feasible directions are symmetric, the gradient of $\bar R$ with respect to $\boldsymbol{\Theta}_\mathrm{T}$ is the symmetric part of~\eqref{eq:DT}, i.e.,
\begin{equation}
\nabla_{\boldsymbol{\Theta}_\mathrm{T}}\bar R=\tfrac12\big(\mathbf{D}_\mathrm{T}+\mathbf{D}_\mathrm{T}^\T\big).
\label{eq:grad-ThetaT}
\end{equation}
The detailed derivations of the gradients~\eqref{eq:grad-ThetaT}, \eqref{eq:grad-ThetaR}, and~\eqref{eq:grad-p} are given in Appendix~\ref{app:gradient}. The projection $\mathcal{T}_{\boldsymbol{\Theta}_\mathrm{T}}(\cdot)$ onto the constraint~\eqref{eq:rp-net} is given by the following lemma.

\begin{lemma}
\label{lem:projection}
Let $\mathbf{A}\in\mathbb{C}^{M\times M}$ be a general matrix whose symmetrization $\frac12(\mathbf{A}+\mathbf{A}^\T)$ is invertible, and let $\frac12(\mathbf{A}+\mathbf{A}^\T)=\mathbf{U}_\mathbf{A}\boldsymbol{\Sigma}_\mathbf{A}\mathbf{V}_\mathbf{A}^\Herm$ be its \ac{SVD}. Then, the Euclidean projection of $\mathbf{A}$ onto the set of $M\times M$ symmetric unitary matrices is unique and given by $\mathcal{T}_{\boldsymbol{\Theta}}(\mathbf{A})=\mathbf{U}_\mathbf{A}\mathbf{V}_\mathbf{A}^\Herm$, so that $\mathcal{T}_{\boldsymbol{\Theta}_\mathrm{T}}(\cdot)=\mathcal{T}_{\boldsymbol{\Theta}_\mathrm{R}}(\cdot)=\mathcal{T}_{\boldsymbol{\Theta}}(\cdot)$.
\end{lemma}

\begin{IEEEproof}
See Appendix~\ref{app:projection}.
\end{IEEEproof}

\subsubsection{Receive Scattering Matrix}
Similarly, $\bar R$ depends on $\boldsymbol{\Theta}_\mathrm{R}$ through three paths: (1) the combiner $\mathbf{G}$, which determines the nominal effective channel $\mathbf{E}$, the nominal filtered-noise power, and the matrix $\mathbf{C}_\mathrm{T}$ of the transmit error term; (2) the matrix $\Phib_\mathrm{R}^\Herm$, shared by the receive and noise error terms; and (3) the matrix $\boldsymbol{\Pi}_\mathrm{R}$ of the same terms. Collecting path (1) into
\begin{equation}
\dot{\mathbf{G}}\triangleq\big(\boldsymbol{\Upsilon}\odot\mathbf{E}\big)(\mathbf{H}\mathbf{F})^\Herm
+\diag(\boldsymbol{\kappa})\mathbf{G}
+\dot{\mathbf{A}}_\mathrm{T}\big(\mathbf{H}\boldsymbol{\Pi}_\mathrm{T}^\T\big)^\Herm,
\label{eq:G-dot}
\end{equation}
and propagating all paths to $\Wb_\mathrm{R}$ through~\eqref{eq:maps-Theta}, the gradient is
\begin{equation}
\nabla_{\boldsymbol{\Theta}_\mathrm{R}}\bar R=\tfrac12\big(\mathbf{D}_\mathrm{R}+\mathbf{D}_\mathrm{R}^\T\big),
\label{eq:grad-ThetaR}
\end{equation}
where $\mathbf{D}_\mathrm{R}\triangleq \mathbf{S}_{\mathrm{R},1}\dot{\mathbf{G}}\mathbf{S}_{\mathrm{R},2}^\T +2\mathbf{S}_{\mathrm{R},1}\big(\dot{\mathbf{A}}_\mathrm{R}+\dot{\mathbf{A}}_\mathrm{N}\big)+2\big(\dot{\mathbf{B}}_\mathrm{R}(\mathbf{H}\mathbf{F})^\Herm+\dot{\mathbf{B}}_\mathrm{N}\big)\mathbf{S}_{\mathrm{R},2}^\T$, and its projection is given by Lemma~\ref{lem:projection}.

\subsubsection{Power Allocation}
The expected rate $\bar R$ depends on $\mathbf{p}$ only through $T_k$ and $J_k$. Differentiating~\eqref{eq:Rbar-TJ} with respect to $p_j$ gives
\begin{equation}
\big[\nabla_{\mathbf{p}}\bar R\big]_j=\frac{P_\mathrm{T}}{\ln2}\left(\sum_{k=1}^{N_\mathrm{S}}\frac{\bar{E}_{k,j}}{T_k}-\sum_{k\ne j}\frac{\bar{E}_{k,j}}{J_k}\right).
\label{eq:grad-p}
\end{equation}
The projection onto the constraints~\eqref{eq:rp-pos} and~\eqref{eq:rp-sum} is $\mathcal{T}_{\mathbf{p}}(\mathbf{a})=[\mathbf{a}-\tau\mathbf{1}]^+$, where $\tau$ ensures that the entries sum to one and is found by sorting the entries of $\mathbf{a}$~\cite{duchi2008efficient}.

\subsubsection{Step Sizes}
\label{subsubsec:algorithm}
The step sizes $\alpha_\mathrm{y}^{(t)}$ and $\alpha_\mathrm{x}^{(t)}$ are chosen by backtracking. Given a point $\mathbf{w}\triangleq(\breve{\boldsymbol{\Theta}}_\mathrm{T},\breve{\boldsymbol{\Theta}}_\mathrm{R},\breve{\mathbf{p}})$ and a trial step size $\alpha$, let $\hat{\mathbf{w}}\triangleq(\hat{\boldsymbol{\Theta}}_\mathrm{T},\hat{\boldsymbol{\Theta}}_\mathrm{R},\hat{\mathbf{p}})$ be the triple obtained by one update and one projection of each variable, and let $\varrho\triangleq\sum_{\boldsymbol{\vartheta}}\mathrm{Re}\{\tr\{(\nabla_{\boldsymbol{\vartheta}}\bar R)^\Herm(\hat{\boldsymbol{\vartheta}}-\breve{\boldsymbol{\vartheta}})\}\}$ and $d^2\triangleq\sum_{\boldsymbol{\vartheta}}\|\hat{\boldsymbol{\vartheta}}-\breve{\boldsymbol{\vartheta}}\|_\mathrm{F}^2$ denote the first-order predicted increase of $\bar R$ and the squared distance moved, where the sums are over the three variables and the gradients are evaluated at $\mathbf{w}$. The trial step size is accepted if
\begin{equation}
\bar R(\hat{\mathbf{w}})\ge\bar R(\mathbf{w})+\varrho-\frac{d^2}{4\alpha},
\label{eq:backtracking}
\end{equation}
and is otherwise reduced to $\eta\alpha$, with $\eta\in(0,1)$. To avoid overly small steps, each backtracking starts from the step size accepted at the previous iteration divided by $\eta$, capped at $\alpha_{\max}$.

Algorithm~\ref{alg:mapg} summarizes the proposed robust design.

\begin{algorithm}[t]
\caption{Robust Design Using Monotone Accelerated Projected Gradient Ascent}
\label{alg:mapg}
\footnotesize
\begin{algorithmic}[1]
\Require \algmlin{$\mathbf{x}^{(0)}=\big(\boldsymbol{\Theta}_\mathrm{T}^{(0)},\boldsymbol{\Theta}_\mathrm{R}^{(0)},\mathbf{p}^{(0)}\big)$, $\alpha_{\max}>0$, $\eta\in(0,1)$, $\epsilon>0$.}
\State \algml{$\mathbf{x}^{(1)}=\mathbf{z}^{(1)}=\mathbf{x}^{(0)}$, $\varphi_0=0$, $\varphi_1=1$, $\alpha_\mathrm{y}=\alpha_\mathrm{x}=\alpha_{\max}$, $t=1$.}
\Repeat
    \State \algml{Compute $\mathbf{y}^{(t)}$ via~\eqref{eq:generic-extrapolation} and the gradients at $\mathbf{y}^{(t)}$ via~\eqref{eq:grad-ThetaT},~\eqref{eq:grad-ThetaR}, and~\eqref{eq:grad-p}, and set $\alpha_\mathrm{y}\gets\min\{\alpha_\mathrm{y}/\eta,\alpha_{\max}\}$.}
    \Do \Comment{accelerated candidate}
        \State \algml{Compute $\mathbf{z}^{(t+1)}$ via~\eqref{eq:generic-update-acc} and~\eqref{eq:generic-projection-acc}.}
        \If{\eqref{eq:backtracking} fails for $\mathbf{y}^{(t)}$ and $\mathbf{z}^{(t+1)}$}
            \State Update $\alpha_\mathrm{y}\gets\eta\,\alpha_\mathrm{y}$.
        \EndIf
    \doWhile{\eqref{eq:backtracking} fails for $\mathbf{y}^{(t)}$ and $\mathbf{z}^{(t+1)}$}
    \State \algml{Compute the gradients at $\mathbf{x}^{(t)}$ and set $\alpha_\mathrm{x}\gets\min\{\alpha_\mathrm{x}/\eta,\alpha_{\max}\}$.}
    \Do \Comment{safeguard candidate}
        \State \algml{Compute $\mathbf{v}^{(t+1)}$ via~\eqref{eq:generic-update-pg} and~\eqref{eq:generic-projection-pg}.}
        \If{\eqref{eq:backtracking} fails for $\mathbf{x}^{(t)}$ and $\mathbf{v}^{(t+1)}$}
            \State Update $\alpha_\mathrm{x}\gets\eta\,\alpha_\mathrm{x}$.
        \EndIf
    \doWhile{\eqref{eq:backtracking} fails for $\mathbf{x}^{(t)}$ and $\mathbf{v}^{(t+1)}$}
    \State \algml{Update $\mathbf{x}^{(t+1)}$ via~\eqref{eq:selection}, $\varphi_{t+1}\gets\frac12\big(1+\sqrt{1+4\varphi_t^2}\big)$, and $t\gets t+1$.}
\Until{$\bar R(\mathbf{x}^{(t)})-\bar R(\mathbf{x}^{(t-1)})\le\epsilon\,\bar R(\mathbf{x}^{(t-1)})$}
\Ensure $\mathbf{x}^{(t)}=\big(\boldsymbol{\Theta}_\mathrm{T}^{(t)},\boldsymbol{\Theta}_\mathrm{R}^{(t)},\mathbf{p}^{(t)}\big)$.
\end{algorithmic}
\end{algorithm}

\vspace{-0.3cm}
\subsection{Convergence Analysis}
\label{subsec:robust-convergence}

The convergence of Algorithm~\ref{alg:mapg} follows from two properties. First, every iterate is the output of a projection, so that all iterates are feasible. Second, the safeguard candidate satisfies~\eqref{eq:backtracking} and, being the projection of a gradient step taken from the feasible point $\mathbf{x}^{(t)}$, satisfies $\varrho\ge d^2/(2\alpha_\mathrm{x})$, which together ensure that $\bar R(\mathbf{v}^{(t+1)})\ge\bar R(\mathbf{x}^{(t)})$, and~\eqref{eq:selection} then gives $\bar R(\mathbf{x}^{(t+1)})\ge\bar R(\mathbf{v}^{(t+1)})$. Hence, $\bar R(\mathbf{x}^{(t)})$ is non-decreasing and, since $\bar R$ is continuous on the compact feasible set and thus bounded, it converges. Moreover, following~\cite{li2015accelerated}, every accumulation point of $\{\mathbf{x}^{(t)}\}$ is a stationary point of~\eqref{eq:robust-problem}. Since the algorithm is initialized with the capacity-achieving configuration and is monotone, the robust design never performs worse than the capacity-achieving design.

\vspace{-0.3cm}
\subsection{Complexity Analysis}
\label{subsec:robust-complexity}

Each iteration of Algorithm~\ref{alg:mapg} requires two gradient evaluations, two updates and projections of each variable, and the evaluations of $\bar R$ in the backtracking. Since the matrix products in~\eqref{eq:omega-final} and~\eqref{eq:backprop} have one dimension equal to $N_\mathrm{S}$, updating $\boldsymbol{\Theta}_\mathrm{T}$, $\boldsymbol{\Theta}_\mathrm{R}$, and $\mathbf{p}$ requires $\mathcal{O}(N_\mathrm{S}M_\mathrm{T}^2+N_\mathrm{S}N_\mathrm{R}M_\mathrm{T})$, $\mathcal{O}(N_\mathrm{S}M_\mathrm{R}^2+N_\mathrm{S}N_\mathrm{T}M_\mathrm{R})$, and $\mathcal{O}(N_\mathrm{S}^2)$ operations, whereas projecting them requires $\mathcal{O}(M_\mathrm{T}^3)$, $\mathcal{O}(M_\mathrm{R}^3)$, and $\mathcal{O}(N_\mathrm{S}\log N_\mathrm{S})$ operations, respectively. Since $N_\mathrm{S}\le\min(N_\mathrm{T},N_\mathrm{R})$, the \acp{SVD} of the projections dominate, and the complexity per iteration is $\mathcal{O}\big(M_\mathrm{T}^3+M_\mathrm{R}^3\big)$.

\vspace{-0.3cm}

\section{Numerical Simulations and Discussions}
\label{sec:results}

In this section, we provide extensive numerical simulations to validate the rate analysis and evaluate the improvement achieved by the robust design, and to compare the \ac{MiLAC} with \ac{DBF} under different hardware impairments. Unless otherwise stated, we use the following parameters. Both the transmitter and receiver employ uniform linear arrays with half-wavelength spacing, where \(N = N_\mathrm{T}=N_\mathrm{R}= 32\), and \( N_\mathrm{S} =8\) streams are transmitted. The reference impedance is \(Z_0=\unit[50]{\Omega}\). The carrier frequency is \(f_\mathrm{c}=\unit[28]{GHz}\), the transmit power budget is \(P_\mathrm{T}=\unit[25]{dBm}\), and the noise power is \(\varpi^2=\unit[-90]{dBm}\). The channel follows Rician fading with a Rician factor of \(K=\unit[3]{dB}\) and a path loss of \(\unit[61.4]{dB}\) at a reference distance of \(\unit[1]{m}\), a path-loss exponent of \(2.4\), and a link distance of \(\unit[50]{m}\). The normalized tolerance is \(\bar{\delta}\in\{0.05,0.1\}\), and the same tolerance applies to both \acp{MiLAC}. Algorithm~\ref{alg:mapg} is run with $\alpha_{\max}=10$, $\eta=0.5$, and $\epsilon=10^{-6}$, for at most $250$ iterations. We average the results over \(100\) independent random channel realizations.

We compare three designs: (i) the \emph{error-free design}, i.e., the capacity-achieving configuration in~\eqref{eq:FGopt} without errors; (ii) the \emph{capacity-achieving design under mismatch}, whose expected rate is evaluated analytically using~\eqref{eq:gammabar-int} and through \ac{MC} simulations, averaging the rate over \(25\) independent error realizations per channel realization; and (iii) the \emph{robust design} obtained using Algorithm~\ref{alg:mapg}, whose analytical rate is the objective $\bar R$ maximized by the algorithm and whose \ac{MC} rate is evaluated as for the second design.

\begin{figure*}[t]
\centering
\begin{minipage}[t]{0.32\textwidth}
\centering
\includegraphics[width=\linewidth]{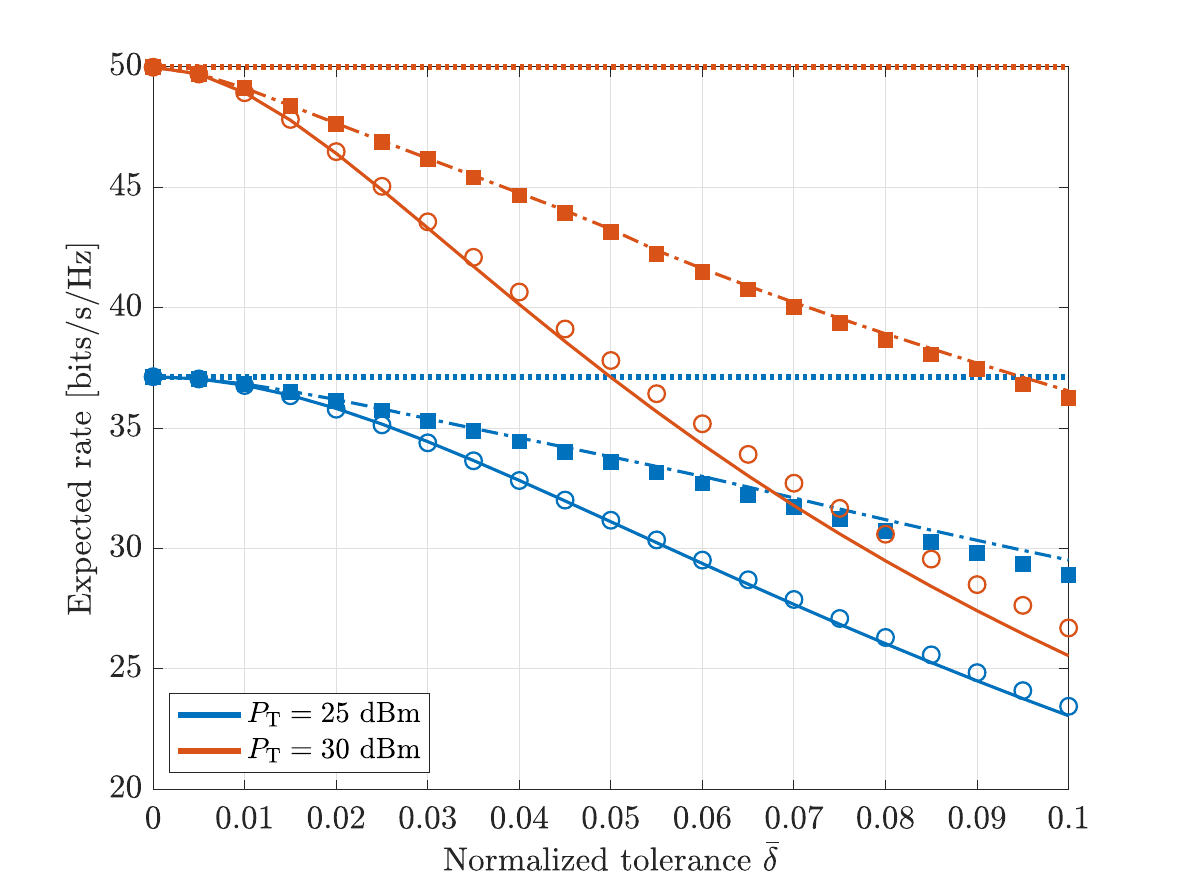}
\vspace{-0.6cm}
\caption{Expected rate versus the normalized tolerance $\bar{\delta}$ for $P_{\mathrm{T}}\in\{25,30\}$~dBm. Dotted lines: error-free rate. Solid lines and circles: capacity-achieving design under mismatch, analytical and simulated. Dash-dot lines and filled squares: robust design, analytical and simulated.}
\label{fig:rate-vs-error}
\end{minipage}\hfill
\begin{minipage}[t]{0.32\textwidth}
\centering
\includegraphics[width=\linewidth]{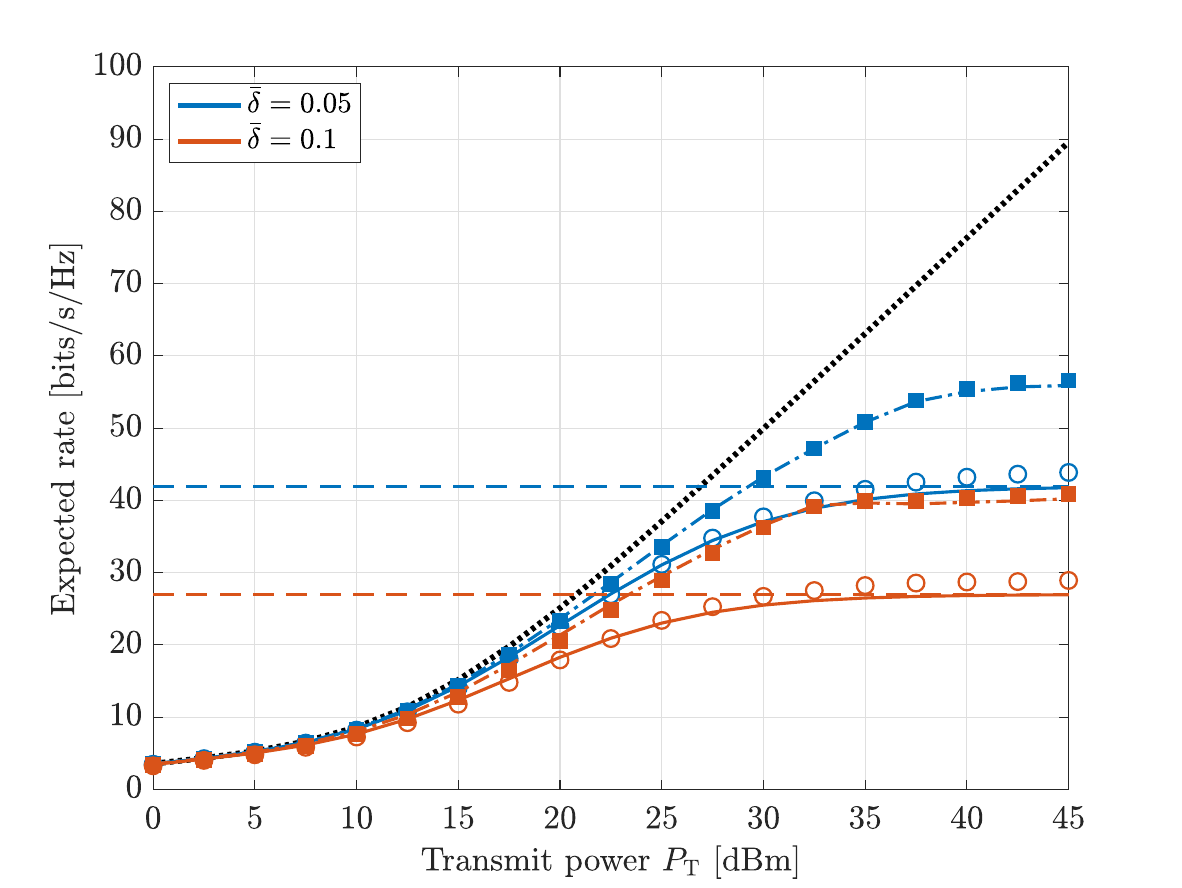}
\vspace{-0.6cm}
\caption{Expected rate versus the transmit power $P_{\mathrm{T}}$ for $\bar{\delta}\in\{0.05,0.1\}$. Dotted lines: error-free rate. Solid lines and circles: capacity-achieving design under mismatch, analytical and simulated. Dash-dot lines and filled squares: robust design, analytical and simulated. Dashed lines: rate ceiling $R_{\mathrm{P}}^{\infty}$ of the capacity-achieving design under mismatch.}
\label{fig:rate-vs-power}
\end{minipage}\hfill
\begin{minipage}[t]{0.32\textwidth}
\centering
\includegraphics[width=\linewidth]{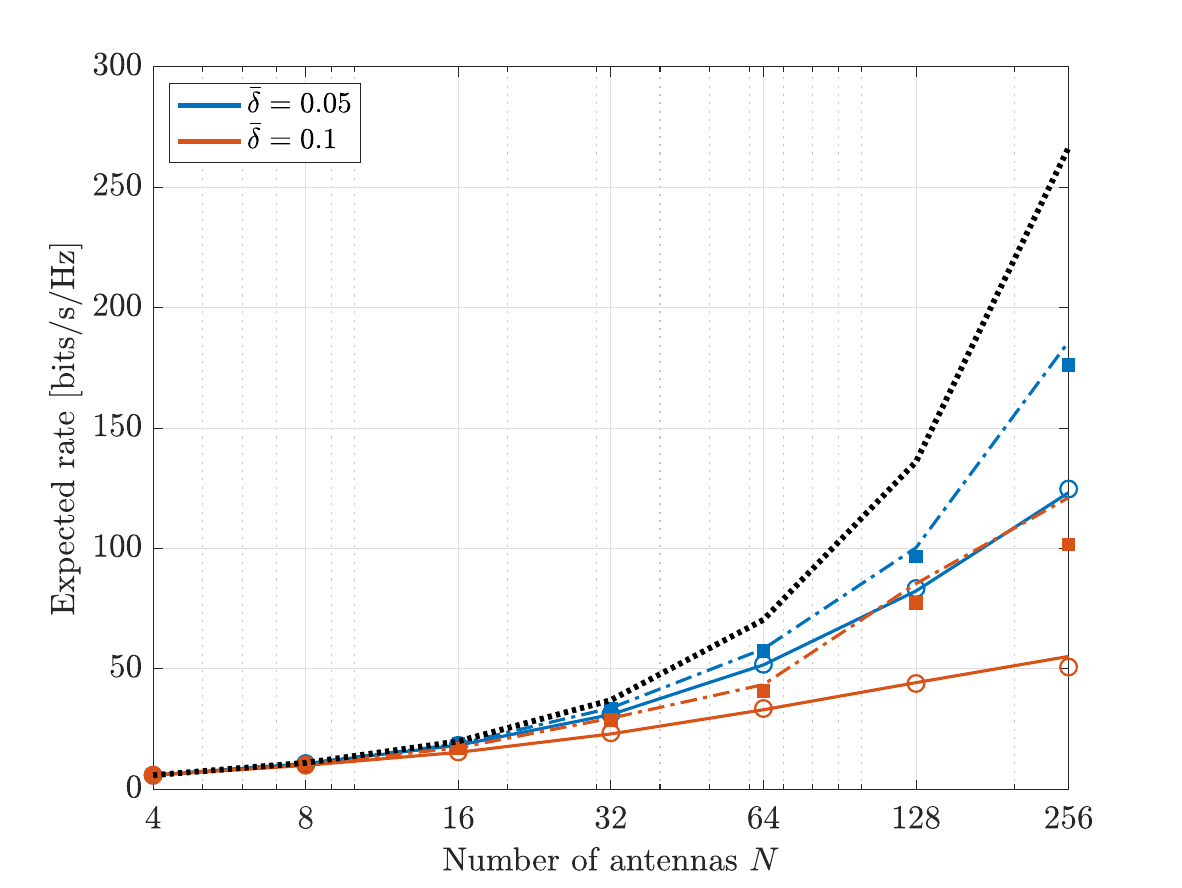}
\vspace{-0.6cm}
\caption{Expected rate versus the number of antennas $N=N_{\mathrm{T}}=N_{\mathrm{R}}$ for $\bar{\delta}\in\{0.05,0.1\}$ and $N_\ell=N_{\mathrm{S}}/N=0.25$. Dotted lines: error-free rate. Solid lines and circles: capacity-achieving design under mismatch, analytical and simulated. Dash-dot lines and filled squares: robust design, analytical and simulated.}
\label{fig:rate-vs-N}
\end{minipage}
\vspace{-0.3cm}
\end{figure*}

\vspace{-0.4cm}
\subsection{Impact of the Component Tolerance}
\label{subsec:results-tolerance}

Fig.~\ref{fig:rate-vs-error} shows the expected rate versus the normalized tolerance $\bar{\delta}$ for two transmit powers. The analytical rate closely follows the simulated one, which validates the first-order analysis, and it is slightly conservative for tolerances above about $0.05$, since the first-order expansion becomes less accurate as the tolerance grows. The rate degradation is substantial even for moderate tolerances. At $\bar{\delta}=0.05$, the mismatch reduces the rate by about $16\%$ at $P_{\mathrm{T}}=25$~dBm and $25\%$ at $P_{\mathrm{T}}=30$~dBm, and, at $\bar{\delta}=0.1$, by about $38\%$ and $49\%$, respectively. The degradation is larger at the higher power, since $\beta_k$ grows with $P_{\mathrm{T}}$, whereas $\chi_k$ does not. Halving the tolerance from $\bar{\delta}=0.1$ to $0.05$ increases the rate by about $8$ and $11.6$~bits/s/Hz at $P_{\mathrm{T}}=25$ and $30$~dBm, respectively, and halving it further to $0.025$ increases it by about $4$ and $7.8$~bits/s/Hz. These gains remain below the limit of $2N_{\mathrm{S}}=16$~bits/s/Hz in Section~\ref{subsec:ins-tolerance}, since the \ac{SINR} improvement factor $(1+\beta_k+\chi_k)/\big(1+(\beta_k+\chi_k)/4\big)$ approaches four only when the interference dominates. Accordingly, the gains are larger at the higher transmit power, where $\beta_k$ is larger, and smaller for tighter tolerances, where the mismatch factors are already small. The robust design recovers a large part of this degradation. At $\bar{\delta}=0.1$, it reduces the degradation from about $38\%$ to $21\%$ at $P_{\mathrm{T}}=25$~dBm and from about $49\%$ to $27\%$ at $P_{\mathrm{T}}=30$~dBm, i.e., it recovers about $44\%$ of the degradation at $P_{\mathrm{T}}=30$~dBm, where it achieves a rate close to the error-free rate at $P_{\mathrm{T}}=25$~dBm.

\vspace{-0.4cm}
\subsection{Impact of the Transmit Power}
\label{subsec:results-power}

Fig.~\ref{fig:rate-vs-power} shows the expected rate versus the transmit power $P_{\mathrm{T}}$ for two tolerances. Without errors, the rate grows without bound as $P_{\mathrm{T}}$ increases. As $P_{\mathrm{T}}$ increases, the interference grows proportionally to $P_{\mathrm{T}}$, so that the rate of the capacity-achieving design under mismatch departs from the error-free rate, earlier for the larger tolerance, and saturates at the ceiling $R_{\mathrm{P}}^{\infty}$, i.e., at about $42$ and $27$~bits/s/Hz for $\bar{\delta}=0.05$ and $0.1$, respectively. Beyond about $35$~dBm, increasing the transmit power brings almost no gain. Halving the tolerance raises the ceiling by about $15$~bits/s/Hz, close to the limit of $2N_{\mathrm{S}}$~bits/s/Hz in Section~\ref{subsec:ins-tolerance}, since the \ac{SINR} improvement approaches a factor of four once the interference dominates. The robust design also saturates, since the errors cannot be removed by design, but about $14$~bits/s/Hz higher, i.e., at about $56$ and $41$~bits/s/Hz, since it accounts for the leakage. Notably, at high power, the robust design with $\bar{\delta}=0.1$ achieves nearly the same rate as the capacity-achieving design under mismatch with $\bar{\delta}=0.05$, i.e., accounting for the errors is about as effective as halving the tolerance.

\vspace{-0.4cm}
\subsection{Impact of the Number of Antennas}
\label{subsec:results-antennas}

Fig.~\ref{fig:rate-vs-N} shows the expected rate versus the number of antennas with $N_\mathrm{S}=N/4$ streams. Without errors, the rate grows linearly with $N$, since both the number of streams and the array gain grow with $N$, and it becomes large for large arrays. Under mismatch, the growth slows down as $N$ increases, since the degradation of each stream grows with $N$. For $\bar{\delta}=0.1$, doubling $N$ from $128$ to $256$ increases the rate of the capacity-achieving design under mismatch by only about $25\%$, and, at $N=256$, the mismatch reduces the rate by about $53\%$ for $\bar{\delta}=0.05$ and $79\%$ for $\bar{\delta}=0.1$. As discussed in Section~\ref{subsec:ins-array}, the relative degradation increases with $N$, but more slowly as it approaches one. For example, for $\bar{\delta}=0.1$, it increases from about $53\%$ to $67\%$ and $79\%$ as $N$ doubles from $64$ to $128$ and $256$, i.e., by less than $N_{\mathrm{S}}/R_{\mathrm{ideal}}\approx 24\%$ per doubling. Moreover, since the transmit power and the loading factor are fixed, quadrupling $N$ has the same impact as doubling $\bar{\delta}$. For example, the relative degradation is about $53\%$ both for $\bar{\delta}=0.1$ at $N=64$ and for $\bar{\delta}=0.05$ at $N=256$, and about $38\%$ for $\bar{\delta}=0.1$ at $N=32$ and $39\%$ for $\bar{\delta}=0.05$ at $N=128$. Hence, larger arrays require tighter component tolerances. The robust design significantly improves this scaling. At $N=256$, it achieves about $176$~bits/s/Hz for $\bar{\delta}=0.05$, i.e., about $2.75$~bits/s/Hz per stream, and $102$~bits/s/Hz for $\bar{\delta}=0.1$, i.e., about $42\%$ and $85\%$ more than the capacity-achieving design under mismatch, respectively. Its gain grows with both $N$ and $\bar{\delta}$, so that accounting for the mismatch is most beneficial for large arrays and loose component tolerances.

\vspace{-0.4cm}
\subsection{Impact of the Number of Streams}
\label{subsec:results-loading}

\begin{figure*}[t]
\centering
\begin{minipage}[t]{0.32\textwidth}
\centering
\includegraphics[width=\linewidth]{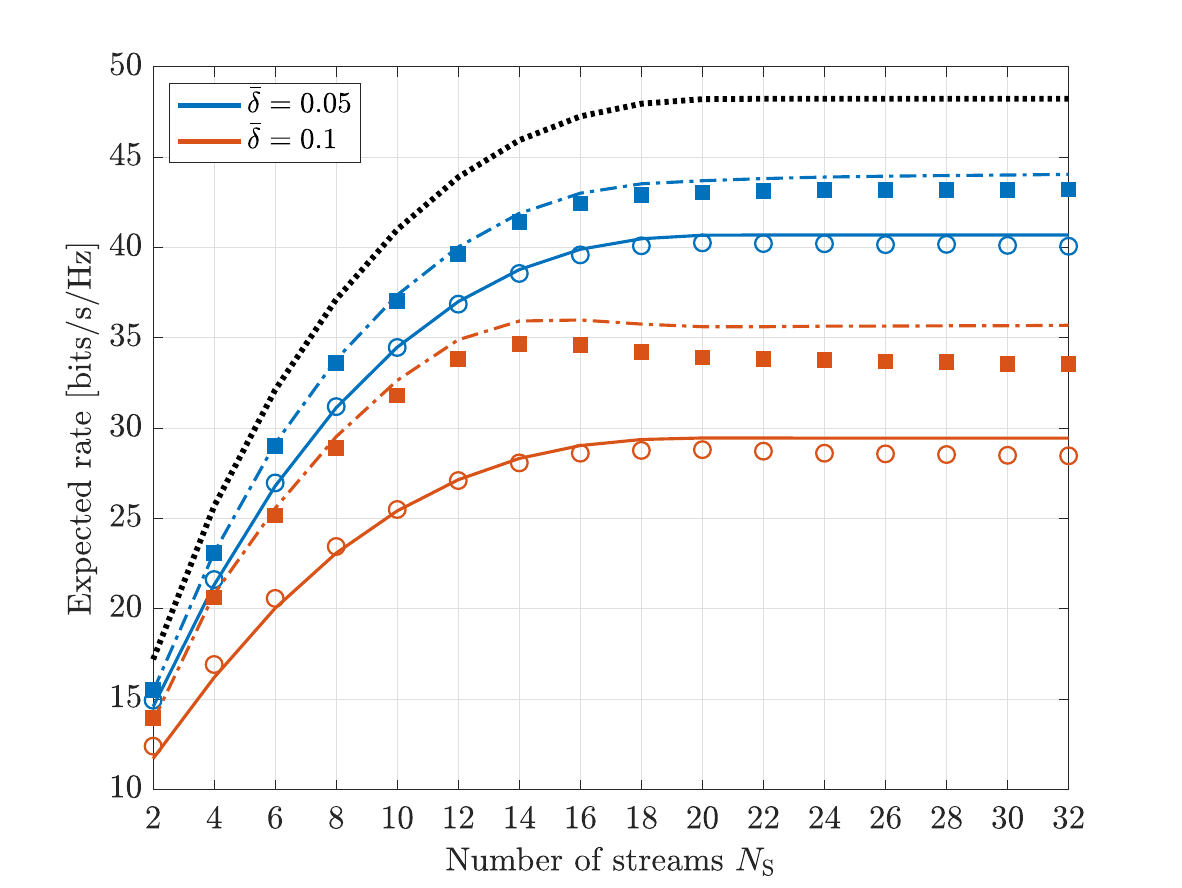}
\vspace{-0.6cm}
\caption{Expected rate versus the number of streams $N_{\mathrm{S}}$ for $\bar{\delta}\in\{0.05,0.1\}$ and $N=N_{\mathrm{T}}=N_{\mathrm{R}}=32$. Dotted lines: error-free rate. Solid lines and circles: capacity-achieving design under mismatch, analytical and simulated. Dash-dot lines and filled squares: robust design, analytical and simulated.}
\label{fig:rate-NS}
\end{minipage}\hfill
\begin{minipage}[t]{0.32\textwidth}
\centering
\includegraphics[width=\linewidth]{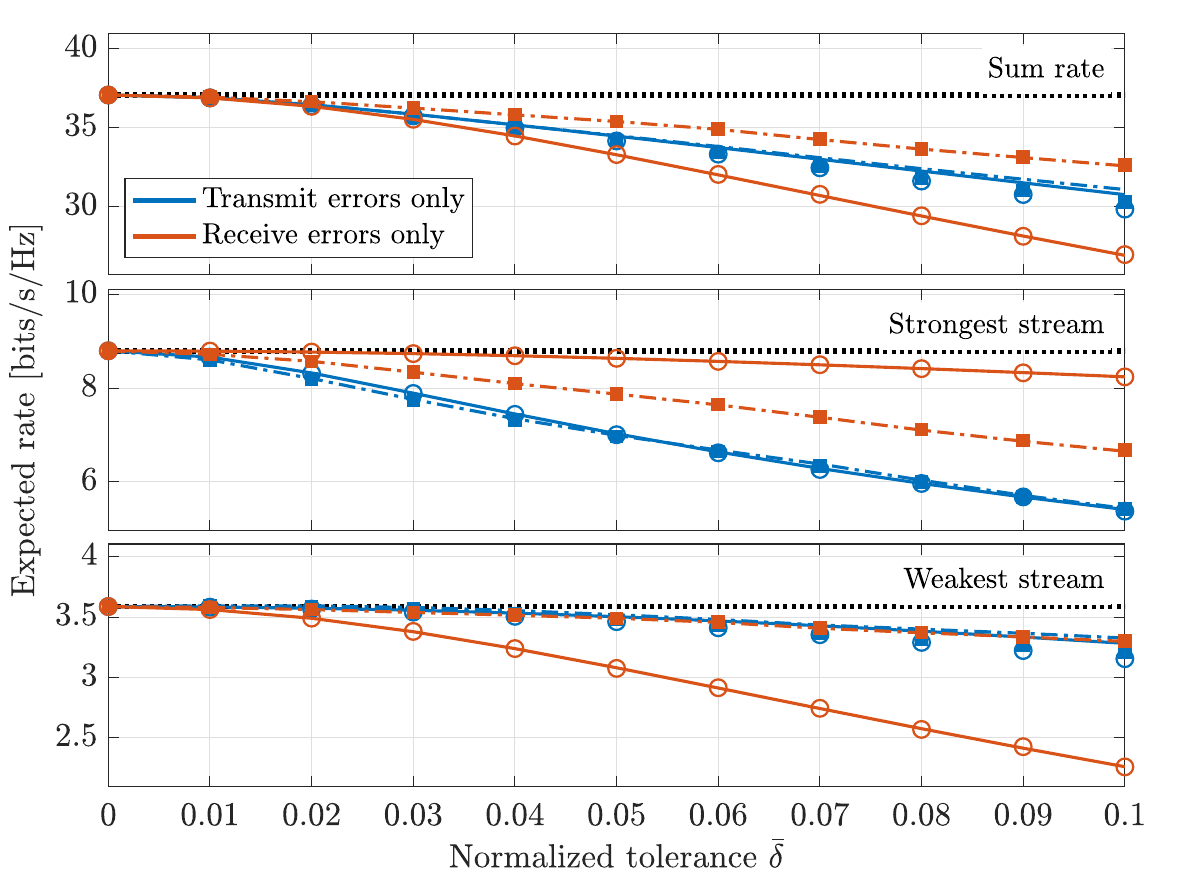}
\vspace{-0.6cm}
\caption{Expected rates versus the normalized tolerance $\bar{\delta}$, with errors in the transmit \ac{MiLAC} only (blue) or in the receive \ac{MiLAC} only (red), while the other \ac{MiLAC} is error-free: sum rate (top), rate of the strongest stream (middle), and rate of the weakest stream (bottom). In all panels, dotted lines: error-free rates; solid lines and circles: capacity-achieving design under mismatch, analytical and simulated; dash-dot lines and filled squares: robust design, analytical and simulated.}
\label{fig:rate-tx-rx}
\end{minipage}\hfill
\begin{minipage}[t]{0.32\textwidth}
\centering
\includegraphics[width=\linewidth]{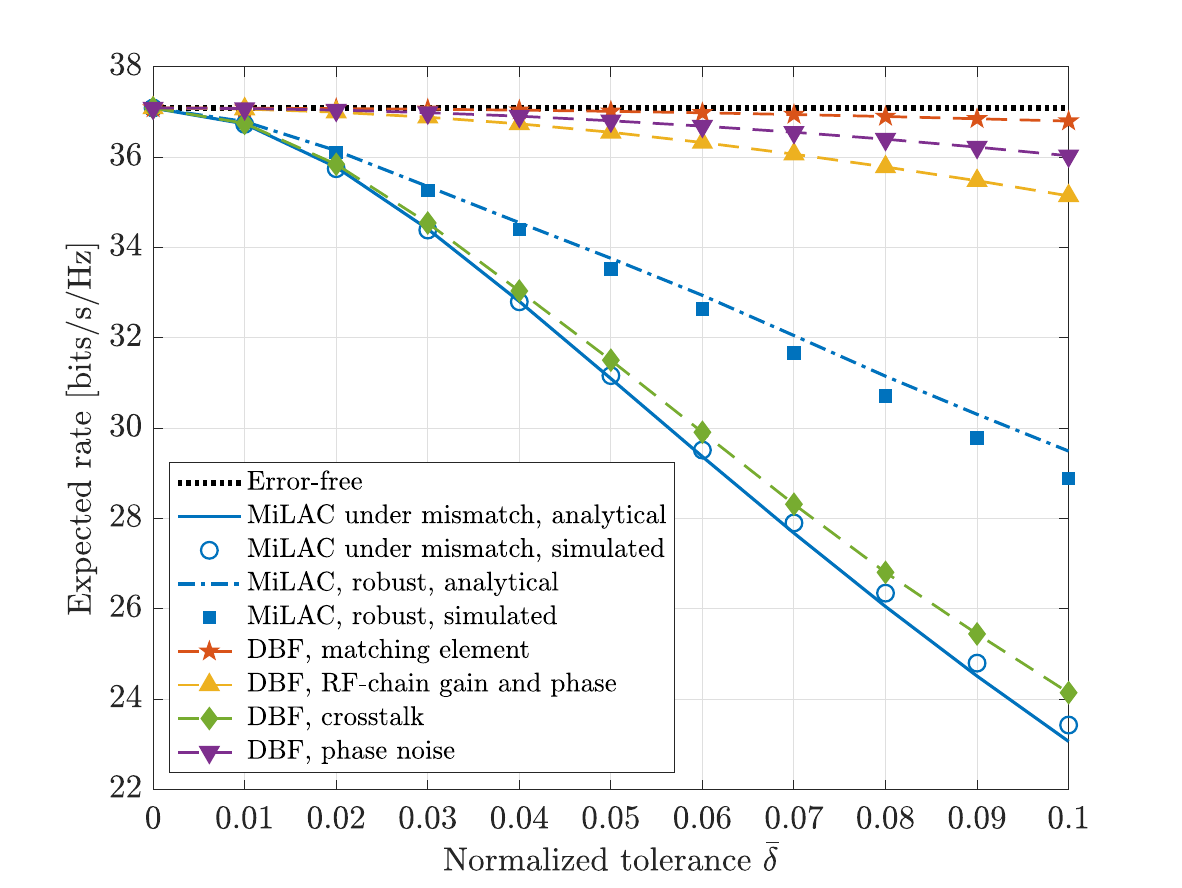}
\vspace{-0.6cm}
\caption{Expected rate versus the normalized tolerance $\bar{\delta}$ for the \ac{MiLAC} under intrinsic impedance mismatch and for \ac{DBF} under four impairments with the same error variance. The error-free rate (dotted line) is the same for both, since both use the capacity-achieving precoder and combiner.}
\label{fig:dbf}
\end{minipage}
\vspace{-0.3cm}
\end{figure*}

Fig.~\ref{fig:rate-NS} shows the expected rate versus the number of streams for $N=32$. All rates increase with $N_{\mathrm{S}}$ up to about $20$ streams and remain constant beyond, since, at $P_{\mathrm{T}}=25$~dBm, water-filling assigns zero power to the weakest eigenmodes. The analytical rate closely follows the simulated one. The relative degradation of the capacity-achieving design under mismatch varies only moderately with $N_{\mathrm{S}}$, i.e., between about $13\%$ and $16\%$ for $\bar{\delta}=0.05$ and between about $32\%$ and $39\%$ for $\bar{\delta}=0.1$. The robust design improves the rate for all $N_{\mathrm{S}}$, by about $8\%$ for $\bar{\delta}=0.05$ and $17\%$ for $\bar{\delta}=0.1$ when $N_{\mathrm{S}}\ge 20$. For $\bar{\delta}=0.1$, its simulated rate peaks at about $14$ streams and slightly decreases beyond, which confirms that a weak stream can cost more rate through its leakage than it contributes. In this range, its analytical rate exceeds the simulated one by about $2$~bits/s/Hz, since the first-order expansion is less accurate for the larger tolerance.

\vspace{-.4cm}
\subsection{Impact of Transmit-Side and Receive-Side Mismatch}
\label{subsec:results-txrx}

Fig.~\ref{fig:rate-tx-rx} isolates the contributions of the two \acp{MiLAC} by applying the errors to one of them only. It can be noted that the receive errors are clearly more detrimental to the sum rate. At $\bar{\delta}=0.1$, they reduce the sum rate by about $27\%$, whereas transmit errors reduce it by about $17\%$. The individual streams reveal the opposite roles of the two \acp{MiLAC}. At $\bar{\delta}=0.1$, transmit errors reduce the rate of the strongest stream by about $39\%$, but that of the weakest stream by only about $6\%$, whereas receive errors reduce the rate of the weakest stream by about $37\%$, but that of the strongest stream by only about $7\%$.

The robust design behaves very differently in the two cases. With receive errors, it reduces the degradation from about $27\%$ to $12\%$ by trading rate of the strongest stream, which decreases from about $8.2$ to $6.6$~bits/s/Hz, for rate of the weaker streams, e.g., the rate of the weakest stream increases from about $2.2$ to $3.2$~bits/s/Hz, close to its error-free value, which increases the sum rate since there are several weak streams and only one dominant stream. With transmit errors, in contrast, it improves the sum rate by only about $0.3$~bits/s/Hz, since the leakage reaches each stream through the same eigenmode as its desired signal and thus cannot be suppressed separately.

\vspace{-0.4cm}
\subsection{Comparison with Digital Beamforming}
\label{subsec:dbf}

\begin{table*}[!t]
\centering
\caption{Intrinsic impedance mismatch of the \ac{MiLAC} and the considered \ac{DBF} impairments, all with error variance $\bar{\varsigma}^2=\bar{\delta}^2/3$. The numbers of error sources correspond to the simulation parameters described in Section~\ref{sec:results}.}
\vspace{-0.3cm}
\label{tab:dbf}
\scriptsize
\renewcommand{\arraystretch}{1.15}
\setlength{\tabcolsep}{3pt}
\begin{tabular}{|>{\raggedright\arraybackslash}p{1.7cm}|>{\raggedright\arraybackslash}p{2.75cm}|>{\raggedright\arraybackslash}p{2.75cm}|>{\raggedright\arraybackslash}p{2.75cm}|>{\raggedright\arraybackslash}p{2.75cm}|>{\raggedright\arraybackslash}p{2.75cm}|}
\hline
 & \textbf{\ac{MiLAC}: impedance mismatch} & \textbf{\ac{DBF}: matching element} & \textbf{\ac{DBF}: \ac{RF}-chain gain and phase} & \textbf{\ac{DBF}: phase noise} & \textbf{\ac{DBF}: crosstalk} \\
\hline
Error source & Reactive components of the network & Reactive element between chain and antenna & Active circuits of each chain & Oscillator of each chain & Coupling between chains \\
\hline
Distribution & Uniform over $[-\bar{\delta},\bar{\delta}]$ & Uniform over $[-\bar{\delta},\bar{\delta}]$ & Gaussian, variance $\bar{\varsigma}^2$ & Gaussian, variance $\bar{\varsigma}^2$, varies per symbol & Magnitude $\bar{\varsigma}$, uniform phase \\
\hline
Number of error sources & $M(M+1)/2=820$  & $N = 32$ & $2N = 64$ & $N = 32$ & $N(N-1) = 992$ \\
\hline
\end{tabular}
\vspace{-0.2cm}
\end{table*}

The \ac{DBF} has no direct counterpart of the intrinsic impedance mismatch, since its weights are computed digitally. To compare it with the \ac{MiLAC}, we therefore consider four \ac{DBF} impairments, applied to the capacity-achieving precoder and combiner, each with the same error variance $\bar{\varsigma}^2$ as the \ac{MiLAC} susceptance errors, and evaluate their expected rates through \ac{MC} simulations.

The first impairment is the tolerance of the reactive matching element between each \ac{RF} chain and its antenna, which has the same tolerance as the \ac{MiLAC} components~\cite{SkyworksSMV1405} and thus provides a like-for-like comparison. A deviation of this element reflects part of the signal and mainly introduces a small phase error in its branch. The second impairment is the gain and phase mismatch of the active circuits of the \ac{RF} chains, such as amplifiers and mixers, which we model as independent zero-mean Gaussian gain and phase errors, as in the per-element perturbation model of~\cite[eq.~(6)]{2019MoghaddamStatistical}. The third impairment is phase noise. Each \ac{RF} chain is driven by its own local oscillator, whose random phase error rotates the signal of its branch and is modeled as Gaussian, as for the Wiener phase process with a separate oscillator per antenna in~\cite[Sec.~2.1]{2015BjornsonDistributed}. The fourth impairment is crosstalk, i.e., part of the signal of each chain leaks into all other chains~\cite{2020HandelMIMO}. As in~\cite[eqs.~(7) and~(15)]{2019HandelImpact}, the chain signals are mixed by a coupling matrix with zero diagonal, whose magnitude is set by the isolation between chains, and we take the coupling phase as random, since it depends on the circuit layout. We consider coupling between all pairs of chains, which mirrors the full connectivity of the \ac{MiLAC}. Table~\ref{tab:dbf} summarizes these impairments.

Fig.~\ref{fig:dbf} compares the expected rate of the \ac{MiLAC} with that of \ac{DBF} under the four impairments. The three impairments that are confined to one antenna branch cause small degradations. At $\bar{\delta}=0.1$, the matching-element tolerance reduces the rate by less than $1\%$, phase noise by about $3\%$, and the \ac{RF}-chain gain and phase mismatch by about $5\%$, whereas the \ac{MiLAC} under mismatch loses about $37\%$ of its error-free rate. The matching element is the least harmful, since its deviation mainly introduces a phase error of half the susceptance error, whereas the \ac{RF}-chain mismatch affects both the gain and the phase. In contrast, crosstalk between all \ac{RF} chains closely follows the \ac{MiLAC}, with a degradation of about $35\%$ at $\bar{\delta}=0.1$. When the errors couple all branches, as in the fully connected \ac{MiLAC} or with crosstalk between all chains, each weight accumulates errors of a fixed size, which are large compared with the small weights of a large array. 

%The robust design with $\bar{\delta}=0.03$ achieves approximately the same rate as \ac{DBF} with \ac{RF}-chain mismatch at $\bar{\delta}=0.1$, so that the \ac{MiLAC} matches this \ac{DBF} benchmark with components whose tolerance is about three times tighter. In return, the \ac{MiLAC} requires only $N_\mathrm{S}$ \ac{RF} chains instead of $N$, and the gain and phase mismatch and the phase noise of these chains only scale or rotate the individual streams, without creating inter-stream interference. Hence, the \ac{MiLAC} trades a higher sensitivity to the tolerance of its passive components for fewer \ac{RF} chains and robustness to the impairments of these chains.

\vspace{-0.3cm}

\section{Conclusion}
\label{sec:conclusion}

This paper analyzed \ac{MiLAC}-aided \ac{MIMO} systems under intrinsic impedance mismatch, modeled by bounded uniform admittance errors, and proposed a robust design that accounts for these errors. Based on a perturbation analysis of the Cayley transform, we derived a closed-form approximation of the expected sum rate, in which the errors appear through a desired-signal perturbation, an inter-stream interference, and a filtered-noise factor, all proportional to the square of the tolerance. The analysis showed that, unlike in the error-free case, the sum rate saturates at high transmit power, that, for a fixed ratio of streams to antennas, doubling the number of antennas is as detrimental as increasing the tolerance by a factor of $\sqrt{2}$, and that errors in the receive \ac{MiLAC} are generally more detrimental than errors in the transmit \ac{MiLAC}. The robust design never performs worse than the capacity-achieving design and, at high power, is about as effective as halving the tolerance. Compared with \ac{DBF}, the \ac{MiLAC} is more sensitive to component tolerances, owing to its full connectivity.

% =====================================================================
\appendices

\vspace{-0.3cm}
\section{Proof of Lemma~\ref{lem:moments-int}}
\label{app:admittance-moments}

The zero mean follows directly from $\Ex[\Delta\Yb]=\mathbf{0}$. Using the symmetry $\Delta Y_{m,n} = \Delta Y_{n,m}$, we decompose $[\mathbf{A}\Delta\Yb\mathbf{B}]_{k,i}$ as
\begin{equation}
[\mathbf{A}\Delta\Yb\mathbf{B}]_{k,i}=\sum_m\Delta Y_{m,m}T_{m,m}+\sum_{m<n}\Delta Y_{m,n}(T_{m,n}+T_{n,m}), \label{eq:bilinear-expansion}
\end{equation}
where $T_{m,n}\triangleq[\mathbf{A}]_{k,m}[\mathbf{B}]_{n,i}$. Since the free entries are independent and zero-mean, with $\Ex[\Delta B_{m,n}^2]=\varsigma^2$, the expectation of its squared magnitude has no cross terms, which gives $\Ex[|[\mathbf{A}\Delta\Yb\mathbf{B}]_{k,i}|^2]=\varsigma^2S_{k,i}$ with $S_{k,i}\triangleq\sum_{m<n}|T_{m,n}+T_{n,m}|^2+\sum_m|T_{m,m}|^2$. Since the summand of the first sum is symmetric in $(m,n)$, the sum over $m<n$ equals half the sum over all $m\ne n$, so that
\begin{equation}
\begin{aligned}
& \sum_{m<n}\big|T_{m,n}+T_{n,m}\big|^2=\ \sum_{m,n}\big|T_{m,n}\big|^2\\
&+\mathrm{Re}\bigg\{\sum_{m,n}T_{m,n}T_{n,m}^*\bigg\} -2\sum_m\big|T_{m,m}\big|^2,
\end{aligned}
\end{equation}
where $\sum_{m,n}|T_{m,n}|^2=\|[\mathbf{A}]_{k,:}\|_2^2\|[\mathbf{B}]_{:,i}\|_2^2$ and $\sum_{m,n}T_{m,n}T_{n,m}^*=\big(\sum_m[\mathbf{A}]_{k,m}[\mathbf{B}]_{m,i}^*\big)\big(\sum_n[\mathbf{A}]_{k,n}^*[\mathbf{B}]_{n,i}\big)=|[\mathbf{A}\mathbf{B}^*]_{k,i}|^2$. Combining these expressions gives $S_{k,i}=[\boldsymbol{\Omega}(\mathbf{A},\mathbf{B})]_{k,i}$ as defined in Lemma~\ref{lem:moments-int}, which proves~\eqref{eq:closed-moments-entry}, and~\eqref{eq:closed-moments-row} follows by summing over $i$.

Finally, since the $\Delta B_{m,n}$ are real, the real and imaginary parts of~\eqref{eq:bilinear-expansion} are weighted sums of the same independent, zero-mean, and bounded variables, so that, by the multivariate Lindeberg central limit theorem, $[\mathbf{A}\Delta\Yb\mathbf{B}]_{k,i}$ approaches a zero-mean complex Gaussian random variable whenever the largest squared weight is negligible compared with the sum of squared weights. \hfill$\blacksquare$

\vspace{-0.3cm}
\section{Proof of Lemma~\ref{lem:completion-average}}
\label{app:completion-average}

Since $\mathbf{Q}$ is symmetric and unitary, the matrix $\boldsymbol{\Psi}_\mathrm{R}=\mathbf{M}^*\mathbf{Q}\mathbf{M}^\Herm$ is symmetric, satisfies
$\boldsymbol{\Psi}_\mathrm{R}\Ub=\mathbf{M}^*\mathbf{Q}\mathbf{M}^\Herm\Ub=\mathbf{0}$, and obeys
$\boldsymbol{\Psi}_\mathrm{R}^\Herm\boldsymbol{\Psi}_\mathrm{R}
=\mathbf{M}\mathbf{Q}^\Herm\mathbf{M}^\T\mathbf{M}^*\mathbf{Q}\mathbf{M}^\Herm
=\mathbf{M}\mathbf{M}^\Herm=\mathbf{I}-\Ub\Ub^\Herm$, using
$\mathbf{M}^\T\mathbf{M}^*=\mathbf{I}$. Conversely, any completion satisfying~\eqref{eq:theta11} annihilates $\range(\Ub)$ and is a partial isometry onto
$\range(\Ub)^\perp$, hence of this form. The set of symmetric unitary matrices is invariant under $\mathbf{Q}\to e^{\mathrm{j}\theta}\mathbf{Q}$ for every $\theta$, so that any distribution on that set invariant under this map, in particular the uniform one, satisfies
$\Ex[\mathbf{Q}]=e^{\mathrm{j}\theta}\Ex[\mathbf{Q}]$ for all $\theta$, so that $\Ex[\mathbf{Q}]=\mathbf{0}$, and $\Ex[\boldsymbol{\Psi}_\mathrm{R}]=\mathbf{0}$ follows by linearity. \hfill$\blacksquare$

\vspace{-0.3cm}
\section{Proof of Lemma~\ref{lem:projection}}
\label{app:projection}

Let $\mathcal{U}_M$ denote the set of $M\times M$ symmetric unitary matrices, and let $\mathbf{S}\triangleq\frac12(\mathbf{A}+\mathbf{A}^\T)$ and $\mathbf{K}\triangleq\frac12(\mathbf{A}-\mathbf{A}^\T)$ denote the symmetric and antisymmetric parts of $\mathbf{A}$. For any $\boldsymbol{\Theta}\in\mathcal{U}_M$, $\|\boldsymbol{\Theta}\|_\mathrm{F}^2=M$, so that $\|\mathbf{A}-\boldsymbol{\Theta}\|_\mathrm{F}^2=\|\mathbf{A}\|_\mathrm{F}^2+M-2\,\mathrm{Re}\{\tr\{\mathbf{A}^\Herm\boldsymbol{\Theta}\}\}$, and the projection maximizes $\mathrm{Re}\{\tr\{\mathbf{A}^\Herm\boldsymbol{\Theta}\}\}$. Using $[\mathbf{K}]_{j,i}=-[\mathbf{K}]_{i,j}$ and $[\boldsymbol{\Theta}]_{j,i}=[\boldsymbol{\Theta}]_{i,j}$ and exchanging the summation indices gives $\tr\{\mathbf{K}^\Herm\boldsymbol{\Theta}\}=\sum_{i,j}[\mathbf{K}]_{j,i}^*[\boldsymbol{\Theta}]_{j,i}=-\sum_{i,j}[\mathbf{K}]_{i,j}^*[\boldsymbol{\Theta}]_{i,j}=0$, so that the antisymmetric part of $\mathbf{A}$ plays no role, and the problem reduces to maximizing $\mathrm{Re}\{\tr\{\mathbf{S}^\Herm\boldsymbol{\Theta}\}\}$ over $\mathcal{U}_M$.

We relax the symmetry constraint and maximize over all unitary matrices. By von Neumann's trace inequality, $|\tr\{\mathbf{S}^\Herm\boldsymbol{\Theta}\}|\le\sum_i\sigma_i(\mathbf{S})$, where $\sigma_i(\mathbf{S})$ is the $i$-th singular value of $\mathbf{S}$, with equality attained by the unitary polar factor $\mathbf{U}_\mathbf{A}\mathbf{V}_\mathbf{A}^\Herm$ of $\mathbf{S}$, which is unique since $\mathbf{S}$ is invertible. Since $\mathbf{S}$ is symmetric, it admits a Takagi factorization $\mathbf{S}=\mathbf{L}\boldsymbol{\Sigma}_\mathbf{A}\mathbf{L}^\T$ with $\mathbf{L}$ unitary, so that $\mathbf{U}_\mathbf{A}=\mathbf{L}$ and $\mathbf{V}_\mathbf{A}=\mathbf{L}^*$ is a valid \ac{SVD}, and $\mathbf{U}_\mathbf{A}\mathbf{V}_\mathbf{A}^\Herm=\mathbf{L}\mathbf{L}^\T$,
which is symmetric and unitary, so that the unique maximizer over the unitary matrices lies in $\mathcal{U}_M$ and is the unique projection onto $\mathcal{U}_M$. Finally, when the singular values of $\mathbf{S}$ are distinct, the factors of any \ac{SVD} of $\mathbf{S}$ differ from $\mathbf{L}$ and $\mathbf{L}^*$ only by a common diagonal unitary matrix, which cancels in $\mathbf{U}_\mathbf{A}\mathbf{V}_\mathbf{A}^\Herm$, so that the projection can be computed from any \ac{SVD}. \hfill$\blacksquare$

\vspace{-0.3cm}
\section{Derivation of the Gradients}
\label{app:gradient}

Consider the weighted sum $\phi_{\boldsymbol{\Gamma}}(\mathbf{A},\mathbf{B})$ defined in Section~\ref{subsec:robust-solution}, with $\boldsymbol{\Gamma}\in\mathbb{R}^{K\times J}$, and let
$\mathbf{X}\triangleq\mathbf{A}\mathbf{B}^*$. Differentiating the three terms of $[\boldsymbol{\Omega}(\mathbf{A},\mathbf{B})]_{k,i}$ with respect to $[\mathbf{A}]_{k,m}^*$, where the second term gives $\partial|[\mathbf{X}]_{k,i}|^2/\partial[\mathbf{A}]_{k,m}^*=[\mathbf{X}]_{k,i}[\mathbf{B}]_{m,i}$ since $\mathbf{X}$ is holomorphic in $\mathbf{A}$, weighting by $[\boldsymbol{\Gamma}]_{k,i}$, and summing over $i$ gives, in matrix form,
\begin{align}
\frac{\partial\phi_{\boldsymbol{\Gamma}}}{\partial\mathbf{A}^*}
=\,&\diag\!\big(\boldsymbol{\Gamma}\mathbf{c}(\mathbf{B})\big)\mathbf{A}
+\big(\boldsymbol{\Gamma}\odot\mathbf{X}\big)\mathbf{B}^\T \nonumber\\
&-\mathbf{A}\odot\Big(\boldsymbol{\Gamma}\big(\mathbf{B}\odot\mathbf{B}^*\big)^\T\Big),
\label{eq:dphi-A}\\
\frac{\partial\phi_{\boldsymbol{\Gamma}}}{\partial\mathbf{B}^*}
=\,&\mathbf{B}\diag\!\big(\boldsymbol{\Gamma}^\T\mathbf{r}(\mathbf{A})\big)
+\mathbf{A}^\T\big(\boldsymbol{\Gamma}\odot\mathbf{X}^*\big) \nonumber\\
&-\mathbf{B}\odot\Big(\big(\mathbf{A}\odot\mathbf{A}^*\big)^\T\boldsymbol{\Gamma}\Big),
\label{eq:dphi-B}
\end{align}
where \eqref{eq:dphi-B} follows by the same steps, noting that $\mathbf{X}$ is anti-holomorphic in
$\mathbf{B}$, so that
$\partial|[\mathbf{X}]_{k,i}|^2/\partial[\mathbf{B}]_{m,i}^*=[\mathbf{X}]_{k,i}^*[\mathbf{A}]_{k,m}$.
Evaluating~\eqref{eq:dphi-A} and~\eqref{eq:dphi-B} at the three pairs gives~\eqref{eq:backprop}.

If a real-valued function $\phi$ depends on a matrix argument through
$\mathbf{A}=\mathbf{M}_1\Wb\mathbf{M}_2$, then the chain rule for Wirtinger derivatives gives
\begin{equation}
\frac{\partial\phi}{\partial\Wb^*}=\mathbf{M}_1^\Herm\frac{\partial\phi}{\partial\mathbf{A}^*}\mathbf{M}_2^\Herm.
\label{eq:chain-holo}
\end{equation}
By~\eqref{eq:FG-Theta} and~\eqref{eq:maps-Theta}, every quantity on which $\bar R$ depends is of this form, i.e., $\mathbf{F}$, $\mathbf{C}_\mathrm{T}$, and $\Phib_\mathrm{T}$ with $\Wb=\Wb_\mathrm{T}$, and $\mathbf{G}$, $\Phib_\mathrm{R}^\Herm$, $\mathbf{C}_\mathrm{R}$, and $\boldsymbol{\Pi}_\mathrm{R}$ with $\Wb=\Wb_\mathrm{R}$, so that no anti-holomorphic path arises and \eqref{eq:chain-holo} applies throughout.

For $\boldsymbol{\Theta}_\mathrm{T}$, the derivative with respect to $\mathbf{F}$ collects the nominal term, $(\mathbf{G}\mathbf{H})^\Herm(\boldsymbol{\Upsilon}\odot\mathbf{E})$, and the receive error term through $\mathbf{C}_\mathrm{R}$, $(\boldsymbol{\Pi}_\mathrm{R}\mathbf{H})^\Herm\dot{\mathbf{B}}_\mathrm{R}$, which gives $\dot{\mathbf{F}}$ in~\eqref{eq:F-dot}. For $\boldsymbol{\Theta}_\mathrm{R}$, the derivative with respect to $\mathbf{G}$ collects the nominal signal term, the nominal filtered-noise term, $\diag(\boldsymbol{\kappa})\mathbf{G}$, and the transmit error term through $\mathbf{C}_\mathrm{T}$, which gives $\dot{\mathbf{G}}$ in~\eqref{eq:G-dot}. Applying~\eqref{eq:chain-holo} to the three paths of each scattering matrix and summing gives $\partial\bar R/\partial\Wb_\mathrm{T}^*=\mathbf{D}_\mathrm{T}$ and $\partial\bar R/\partial\Wb_\mathrm{R}^*=\mathbf{D}_\mathrm{R}$.

Since $\boldsymbol{\Theta}=2\Wb-\mathbf{I}$, we have $\mathrm{d}\bar R=2\,\mathrm{Re}\{\tr\{\mathbf{D}^\Herm\mathrm{d}\Wb\}\}=\mathrm{Re}\{\tr\{\mathbf{D}^\Herm\mathrm{d}\boldsymbol{\Theta}\}\}$. The feasible perturbations $\mathrm{d}\boldsymbol{\Theta}$ are symmetric, and, as in Appendix~\ref{app:projection}, the antisymmetric part of $\mathbf{D}$ is orthogonal to every symmetric matrix under the real inner product $\mathrm{Re}\{\tr\{\cdot^\Herm\cdot\}\}$. Hence, $\mathrm{d}\bar R$ is unchanged when $\mathbf{D}$ is replaced by its symmetric part, which gives~\eqref{eq:grad-ThetaT} and~\eqref{eq:grad-ThetaR}.

Finally, since $\bar{E}_{k,i}$ and $\bar{G}_k$ do not depend on $\mathbf{p}$, \eqref{eq:grad-p} follows from~\eqref{eq:Rbar-TJ} with $\partial T_k/\partial p_j=P_\mathrm{T}\bar{E}_{k,j}$ for every $k$, and $\partial J_k/\partial p_j=P_\mathrm{T}\bar{E}_{k,j}$ for $k\ne j$ and zero for $k=j$. \hfill$\blacksquare$

\bibliographystyle{IEEEtran}
\bibliography{Bibliography}

\end{document}